\documentclass[twocolumn]{aastex63} % anonymous
\pdfoutput=1

\newcommand{{\drp}}{{\tt DRP}}
\newcommand{{\dap}}{{\tt DAP}}

\newcommand{\mileshc}{{\tt MILES-HC}}
\newcommand{\marvin}{{\tt Marvin}}
\newcommand{\hiiexplorer}{{\sc HIIexplorer}}
\newcommand{\ha}{{H$\alpha$}}
\newcommand{\miles}{{\tt MILES}}
\usepackage{CJKutf8}
\received{2020 January 2}
\revised{2020 April 24}
\accepted{2020 May 18}

\submitjournal{ApJ}

\shorttitle{Wolf-Rayet galaxy catalog from MaNGA}
\shortauthors{Liang et al.}

\graphicspath{{./}{figures/}}

\begin{document}

\title{Wolf-Rayet galaxies in SDSS-IV MaNGA. I. Catalog construction and sample properties}

\correspondingauthor{Cheng Li}
\email{cli2015@tsinghua.edu.cn}

%\begin{CJK*}{UTF8}{gbsn}

\author[0000-0003-2496-1247]{\begin{CJK*}{UTF8}{gbsn}
Fu-Heng Liang (梁赋珩)\end{CJK*}
}
\affiliation{Department of Astronomy, Tsinghua University, Beijing 100084, China}

\author[0000-0002-8711-8970]{Cheng Li}
\affiliation{Department of Astronomy, Tsinghua University, Beijing 100084, China}

\author{Niu Li}
\affiliation{Department of Astronomy, Tsinghua University, Beijing 100084, China}

\author{Renbin Yan}
\affiliation{Department of Physics \& Astronomy, University of Kentucky, Lexington, KY 40506, USA}

\author{Houjun Mo}
\affiliation{Department of Astronomy, Tsinghua University, Beijing 100084, China}
\affiliation{Department of Astronomy, University of Massachusetts Amherst, MA 01003, USA}

\author{Wei Zhang}
\affiliation{National Astronomical Observatories, Chinese Academy of Sciences, 20A Datun Road, Chaoyang District, Beijing 100012, China}

\author{Camilo Machuca}
\affiliation{Department of Astronomy,  University of Wisconsin, 475 N. Charter Street, Madison, WI 53706, USA}

\author[0000-0002-1379-4204]{Alexandre Roman-Lopes}
\affiliation{Department of Physics \& Astronomy - Universidad de La Serena - Av. Juan Cisternas, 1200 North, La Serena, Chile - aroman@userena.cl}

%\end{CJK*}

% Warning: do not forget to change the display number when new co-authors are added.

%\collaboration{7}{(MaNGA collaboration)}

\begin{abstract}

Wolf-Rayet (WR) galaxies are a rare population of galaxies that host
living high-mass stars during their WR phase (i.e. WR stars), and are
thus expected to provide interesting constraints on the stellar
Initial Mass Function, massive star formation, stellar evolution
models, etc. Spatially resolved spectroscopy should in principle
provide a more efficient way of identifying WR galaxies than
single-fiber surveys of galactic centers such as SDSS-I \& II, as WR stars
should be more preferentially found in discs.  Using Integral Field
Unit data from the ongoing SDSS-IV MaNGA survey, we have performed a
thorough search for WR galaxies in a two-step method. We first
identify \ion{H}{2} regions in each datacube and carry out full
spectral fitting to the stacked spectra.  We then visually inspect the
residual spectrum of each \ion{H}{2} region and identify WR regions
that present a significant ``blue bump'' at $4600-4750$ {\AA}.  The
resulting WR catalog includes 267 WR regions of $\sim$500 pc (radius)
sizes, distributed in 90 galaxies from the current sample of MaNGA
(MaNGA Product Launch 7). We find  WR regions  are exclusively found
in galaxies that show bluest colors and highest star formation rates
for their mass. Most WR galaxies have late-type
  morphologies and show relatively large asymmetry in their images,
  implying that WR regions are more preferentially found in
  interacting/merging galaxies. We estimate the stellar mass
function of WR galaxies, and the mass-dependent detection rate.  The
detection rate of WR galaxies is typically $\sim$2\%, with weak
dependence on stellar mass. This detection rate is about 40 times
higher than previous studies with SDSS single fiber data, and by a factor
of 2 lower than the CALIFA-based WR catalog. We make comparisons with
SDSS and CALIFA studies, and conclude that different detection rates
of different studies can be explained mainly by three factors: spatial
coverage, spectral signal-to-noise ratio, and redshift ranges of the
parent sample. We tabulate the WR galaxy properties for future
studies.
\end{abstract}


\section{Introduction}

Wolf-Rayet (WR) galaxies are a rare population of galaxies 
showing significant feature of WR stars, which were initially
identified by \citet{Wolf-Rayet-67} and are believed to evolve
from O-type stars with an initial mass of 25 $M_\odot$ or larger.
WR stars manifest their existence by presenting a series of 
broad emission lines in the optical wavelength range such as the broad \ion{He}{2} line at 4686 {\AA},
produced by their dense stellar winds \citep{Crowther-07}.
Due to their small number at birth and short lifetime, WR stars
are expected to be a small fraction of the stellar population
in a galaxy, and therefore WR galaxies must be a rare population
as well.

The Milky Way is expected to have thousands of WR stars, and several
hundred have been detected \citep{vanderHucht-01}
and is presented in a continuously maintained online catalog
\footnote{Milky Way WR star catalog:
  \url{http://pacrowther.staff.shef.ac.uk/WRcat}}.  Extra-galactic WR
features were firstly identified by \citet{Allen-Wright-Goss-76} in
the galaxy He 2-10,  which was later termed ``Wolf-Rayet galaxy'' by
\citet{Osterbrock-Cohen-82}.  By the end of the last century, a total
of only 139 WR galaxies beyond the Local Group were reported
\citep{Schaerer-Contini-Pindao-99}.  Most cases were fortuitous
discoveries, and only a few resulted from  intentional systematic
searches through spectroscopy \citep[e.g.]{Kunth-Joubert-85} or
narrow-band imaging  \citep[e.g.]{Drissen-Roy-Moffat-93}.  Thanks to
the large spectroscopic galaxy sample from the Sloan Digital Sky
Survey \citep[SDSS;][]{York-00}, the detection and study of WR galaxies
have advanced dramatically.  \citet{Zhang-07} published the first
SDSS-based WR galaxy catalog with 174 WR galaxies from the SDSS Data
Release 3 (DR3).  Two more catalogs based on later data releases
consist of 570 \citep[][from SDSS DR6]{Brinchmann-Kunth-Durret-08}
and 271 WR galaxies \citep[][from blue compact dwarf galaxies in SDSS DR7]{Agienko-Guseva-Izotov-13},
respectively.
Despite different selection procedures and criteria adopted
in these studies, the fraction of WR galaxies in the SDSS
samples has consistently been very small, at the order of $\sim$0.05\%.

The low detection rate of WR galaxies should be partially 
attributed to the fact that SDSS spectroscopy is limited
to the central 1-2 kpc of galaxies. It is natural to expect
higher detection rates in outer regions of galaxies
considering that star formation occurs more widely in galactic
discs than in their centers.
In addition, the actural area covered by the three-arcsec
  fiber of SDSS can substantially vary over the (though narrow)
  redshift range of the survey, thus diluting WR signatures
  from the center of distant galaxies due to inclusion of non-WR
  area at large radii, while missing off-center WR signatures
  in nearby galaxies due to the limited aperture of the fiber.
  This aperture bias can be largely overcome using the technique
  of integral field unit (IFU) which obtains spatially resolved
  spectroscopy out to large radii in/around a galaxy.
Indeed, detection of WR galaxies has advanced in recent years thanks
to the recent/ongoing IFU surveys, allowing searches
for WR regions across the whole galaxy. \citet{Miralles-Caballero-16}
has recently applied an automated searching procedure
to the IFU data from the Calar Alto Legacy Integral Field Area survey
\citep[CALIFA;][]{Sanchez-12a}, identifying 44 WR regions
in 25 galaxies out of a total of 558 galaxies at $0.005<z<0.03$.
About one third of the WR regions are located within $\sim$1kpc
from the center of their host galaxies. 
Both the fraction of WR galaxies ($\sim 5\%$)
and the fraction of central WR regions ($\sim 1/3$) in the
CALIFA sample are much higher than the WR galaxy fraction
($\sim0.05\%$) in the SDSS sample, which cannot be simply
explained by the different sample selections of the two surveys
or the limited spatial coverage of the SDSS single-fiber spectroscopy.

WR galaxies are interesting not only for their rareness. They have
been used as unique probes of massive star evolution, ionization
origin of ions (e.g. \ion{He}{2}) and dense stellar winds in galaxies,
thus providing important constraints on stellar population synthesis
models of galaxies.  For instance, by analyzing long-slit spectra of
39 WR galaxies, \citet{Guseva-Izotov-Thuan-00} found the relative
number of WR stars to O stars to decrease with decreasing metallicity, in
agreement with evolutional stellar population synthesis models. In
addition, it was found that galaxies with \ion{He}{2} {$\lambda$}4686
emission do not always present WR features, indicating that WR stars
are not the only ionizing source of \ion{He}{2}. This finding was
confirmed and discussed in  \citet{Brinchmann-Kunth-Durret-08} and
\citet{Shirazi-Brinchmann-12}.  With the first SDSS-based WR catalog,
\citet{Zhang-07} performed a comparison of the WR emission of
galaxies with theoretical predictions from evolutionary synthesis
models following \citet{Guseva-Izotov-Thuan-00}, finding that a metallicity-dependent variation of the slope of
stellar Initial Mass Function (IMF) appears to be necessary in order
for the models to agree with the data. Using a larger sample of WR
galaxies selected from a later SDSS data release,
\citet{Brinchmann-Kunth-Durret-08} found the likelihood of galaxies
showing WR features increases with increasing metallicity, although
the WR galaxies present a wide range in morphology. In particular, WR
galaxies showed an elevated nitrogen-to-oxrygen (N/O) ratio relative
to non-WR galaxies, implying a rapid enrichment of the interstellar
medium (ISM) from WR winds. IFU data available in
recent years have been used to further study the N/O ratio of WR
galaxies as supporting evidence for metal pollution from WR winds
\citep[e.g.][]{Perez-Montero-13b, Miralles-Caballero-14}.  The WR
catalog constructed from CALIFA by \citet{Miralles-Caballero-16} has
revealed the similarity between WR galaxies and Gamma-Ray Burst host
galaxies, as well as the importance of binary stellar evolution for
modeling the WR emission at low metallicity.

In this paper we present a thorough search of WR galaxies in the
ongoing Mapping Nearby Galaxies at Apache Point Observatory
\citep[MaNGA;][]{Bundy-15} survey. As one of the three major
experiments of the fourth generation of the Sloan Digital Sky Survey
\citep[SDSS-IV;][]{Blanton-17}, MaNGA is obtaining IFU data for 10,000
galaxies at $0.01<z<0.15$ selected from the SDSS galaxy sample. We identify
our WR galaxies in a two-step method, in which we firstly identify
\ion{H}{2} regions according to the two-dimensional map of H$\alpha$
surface brigthness of each galaxy, and then visually inspect the
\textit{integrated} spectrum of each \ion{H}{2} region obtained by
stacking the original spectra of all spaxels falling in the
region. Following previous studies, we classify an \ion{H}{2} region to
be a WR region if it presents a significant \textit{blue bump} over
the wavelength range $4600-4750$ {\AA}.  This bump is a blend of broad emission lines from \ion{He}{2},
\ion{N}{3}, \ion{N}{5}, \ion{C}{3} and \ion{C}{4} in stellar winds of
WR stars. The ratios among these broad lines vary with the number
ratio of carbon-rich WR stars (namely WC star) and nitrogen-rich WR
stars (namely WN star). Another signature of WR galaxies is a red bump
around 5800 {\AA} from broad emission lines of \ion{C}{3} and
\ion{C}{4}. Normally the red bump is much fainter than the blue bump
and other WR features are even fainter than the red bump. Therefore,
most searches for WR galaxies including this work have made use of the
blue bump signature. 
Out of the 4621 galaxies from MaNGA Product Lauch 7 (MPL-7), we
have constructed a catalog of 90 WR galaxies including a total of
267 WR regions. In this paper we present the identification process
of these WR regions, as well as the global properties of the sample.
In a parallel paper, we study the spatial %  (Liang et al. in prep)
distribution of the WR regions and dependence on galaxy properties.

The rest of the paper is arranged as
follows. Section \S~\ref{sec:data_and_selection} presents a description of
the SDSS-IV MaNGA data and our searching procedure of WR
galaxies. Section \S~\ref{sec:global} presents the catalog and basic
properties of our WR galaxies including the mass-dependent detection
rate and scaling correlations of mass, color and metallicity.  In
\S~\ref{sec:discussion} we discuss our results and connect them
with the litereatue. We summarize in \S~\ref{sec:conclusion}.

\section{Data and selection procedure}
\label{sec:data_and_selection}
\subsection{Overview of the MaNGA survey}
\label{sec:manga_overview}

As one of the three core surveys of the SDSS-IV project
\citep{Blanton-17}, MaNGA aims to obtain integral-field spectroscopy
for an unprecedented sample of 10,000 nearby galaxies
with $0.01<z<0.15$ over a six-year survey period from
July 2014 through June 2020 \citep{Bundy-15}. MaNGA
utilizes  the two dual-channel BOSS spectrographs
at the 2.5-meter Sloan Telescope
\citep{Gunn-06, Smee-13}, covering a wavelengh range
of 3622-10354 {\AA} with a spectral resolution
R$\sim$2000, and reaching a target $r$-band signal-to-noise
$S/N=4-8$ ({\AA}$^{-1}$ per $2^{\prime\prime}$-fiber) at
1-2 $R_e$ (effective radius) with a typical exposure time
of 3 hours. MaNGA uses 29 fiber bundles to obtain the IFU data, including
12 seven-fiber \textit{mini-bundles} for flux calibration
and 17 science bundles with five different field of views
(FoVs) ranging from 12$^{\prime\prime}$ to 32$^{\prime\prime}$,
covered by different numbers of fibers ranging from 19 up to
127. MaNGA instrumentation is described in detail in \cite{Drory-15}.

MaNGA targets are selected from the NASA Sloan Atlas
v1\_0\_1 (NSA)\footnote{\url{https://www.sdss.org/dr13/manga/manga-target-selection/nsa/}}, a catalog constructed
by \cite{Blanton-11} including physical parameters for
{$\sim$}640,000 galaxies from GALEX, SDSS and 2MASS.
\citet{Wake-17} describe the MaNGA sample selection, which
was designed and optimized so as to simultaneously optimize the
IFU size distribution, the IFU allocation strategy and the number
density of targets. The sample consists of three subsamples:
the Primary and Secondary samples having a flat distribution of
the $K$-corrected $i$-band absolute magnitude ($M_i$) and covering
out to 1.5 and 2.5{$R_e$} respectively. The third subsample,
the Color-Enhanced sample selects galaxies on the plane of
$NUV-i$ color versus $M_i$ that are not well sampled by the
Primary sample. Overall, the MaNGA sample covers the stellar
mass range $5\times10^8M_\odot h^{-2}< M_\ast< 3\times10^{11}M_\odot h^{-2}$
with a median redshift of $z\sim0.03$.

MaNGA raw data are reduced with the Data Reduction Pipeline
\citep[{\drp};][]{Law-16}. The {\drp} product for each galaxy is provided in
the form of a datacube with a spaxel size of {0.5\arcsec}, and the
effective spatial resolution of the datacubes can be described by a
Gaussian with a full width at half maximum (FWHM) $\sim${2.5\arcsec}.
Flux calibration, survey strategy and data quality tests are described in detail
in \cite{Yan-16a, Yan-16b}. For more than 80\% of the wavelength
range of MaNGA, the absolute flux calibration is better than 5\%.

In this work we make use of 
MaNGA Product Launch-7 (MPL-7), which contains
4688 datacubes for 4621 unique galaxies.
The MPL-7 is identical to the MaNGA data included in the
SDSS data release 15 \citep[DR15;][]{Aguado-19}.
In addition to the reduced datacubes from
the {\drp}, the DR15 also provides products of the Data Analysis
Pipeline ({\dap}) developed by MaNGA collaboration
\citep{Westfall-19}. The {\dap} performs full spectral fitting
to the {\drp} datacubes using the {\mileshc} stellar spectral library
\citep{Falcon-Barroso-11}, producing measurements of
kinematic parameters, emission line profiles, stellar indicies, etc.
We use {\drp} spectra and {\dap} products as a starting point in
our precedure of searching for WR regions, as described in
\autoref{procedure} below. We also take advantage of the data
visualization and access tool of MaNGA {\marvin}
%{\sc Marvin}
\footnote{\url{ https://dr15.sdss.org/marvin}} developed by
\cite{Cherinka-19}, which includes an online part enabling 
individual galaxies to be examined quickly and conveniently.

\subsection{Overview of WR Searching procedure}
\label{procedure}

We search for WR galaxie in a two-step scheme. For each galaxy
in MaNGA MPL-7, we firstly identify \ion{H}{2} regions using
the two-dimensional map of H$\alpha$ surface brightness provided
by the {\dap}. For each \ion{H}{2} region, we stack the {\drp}
spectra of all spaxels falling in the region. Next, we perform
full spectral fitting to all the stacked spectra, and we visually
inspect the residual spectra to identify WR regions.
An \ion{H}{2} region is identified to be a WR region if it presents
a significant blue bump in the residual spectrum, and 
a galaxy is identified to be a WR galaxy if it contains one or more
WR regions. In the following subsections we
describe our searching procedure in detail. We focus on the
blue bump WR feature in the searching, and we will discuss other
WR features such as the red bump in later sections.

\subsection{Identification of \ion{H}{2} regions}
\label{hii_region}

We assume that WR stars are found exclusively in star-forming regions
with significant \ion{H}{2} emission. Therefore, we start by
identifying \ion{H}{2} regions from the IFU datacube of each
galaxy in MPL-7.  WR features are then visually identified from the
\textit{total} spectrum of each \ion{H}{2} region, obtained from
stacking the original spectra of all the spaxels in the region. As we
will show, the stacking significantly increases the spectral S/N,
allowing those relatively weak WR feature to be more clearly seen.

We use the \ion{H}{2} region finder called
{\hiiexplorer}\footnote{\url{http://www.astroscu.unam.mx/~sfsanchez/HII_explorer/index.html}}
developed by \citet{Sanchez-12b} for the identification of \ion{H}{2}
regions. The overall workflow of the {\hiiexplorer} can be found in
\citet[figure 1]{Sanchez-12b}. In short, from a given {\drp} datacube,
spaxels with H$\alpha$ surface brightness $\Sigma H\alpha
>\Sigma_{\mbox{peak}}$ are picked up in the first place as the central
\textit{peak} of potential \ion{H}{2} regions.  Next, for each peak
spaxel, the spaxels in the vicinity are appended to the region if they
meet the following three criteria: 1) H$\alpha$ surface brightness is
substantially high with $\Sigma H\alpha > \Sigma_{\mbox{min}}$, 2) the
ratio of its brightness to the central spaxel is substantially high
with $\Sigma H\alpha/\Sigma_{cen} H\alpha > f_{\mbox{min}}$, and 3)
its distance from the central spaxel  does not exceed
$r_p^{\mbox{max}}$. In practise, the algorithm starts with the highest
\textit{peak} with largest $\Sigma H\alpha$, and grows it by appending
adjacent spaxels that meet the above requirements, before moving on to
the highest peak in the remaining map. This process is repeated until
every spaxel in the datacube is assigned to an \ion{H}{2} region, or
rejected.

This searching procedure guarantees every \ion{H}{2}
region is roughly centred at the peak value and is roughly round and
symmetric. The minimum brightness ratio ($f_{min}$)
is required so as to make each region roughly coherent, while
spaxels rejected due to this requirement may still be appended to
other neighbouring regions. 
In our work the three threshold parameters are empirically set to be
$\Sigma_{\mbox{peak}}=10^{39.8}$ erg s$^{-1}$kpc$^{-2}$,
$\Sigma_{\mbox{min}}=10^{39.5}$ erg s$^{-1}$kpc$^{-2}$,
$f_{\mbox{min}}=0.1$, and $r_p^{\mbox{max}}= max$(1.5\arcsec, 500pc).
The maximum distance ($r_p^{\mbox{max}}$)
reflects the trade-off between higher S/N from stacking more
spaxels in each region and stronger dilution effect to WR signatures
with possible inclusion of non-WR spaxels.
The typical size of \ion{H}{2} regions ranges from a few
to hundreds of parsecs \citep{Kennicutt-84, Garay-Lizano-99, Kim-Koo-01,
  Hunt-Hirashita-09, Lopez-11, Anderson-19},
so we cannot resolve individual \ion{H}{2} regions due to the limited
spatial resolution of MaNGA which is $\sim$2.5\arcsec,
corresponding to $\sim$1.5 kpc at the median redshift
of MaNGA ($z=0.03$). The maximum distance adopted above is
comparable to (a half of) the MaNGA spatial resolution.
Therefore, each of the \ion{H}{2} regions
identified from the MaNGA datacubes is actually a mixture of
\ion{H}{2} emission and the surrounding diffuse ionized gas (DIG).
The thresholds in $\Sigma H\alpha$ ensure that the resulting
\ion{H}{2} regions are dominated by real \ion{H}{2} emission.
As \cite{Zhang-17} have shown based on a study of the DIG in
MaNGA galaxies, $\Sigma H\alpha$ can be used to effectively
separate \ion{H}{2}-dominated regions from DIG-dominated regions.

\begin{figure}
	\includegraphics[width=\columnwidth]{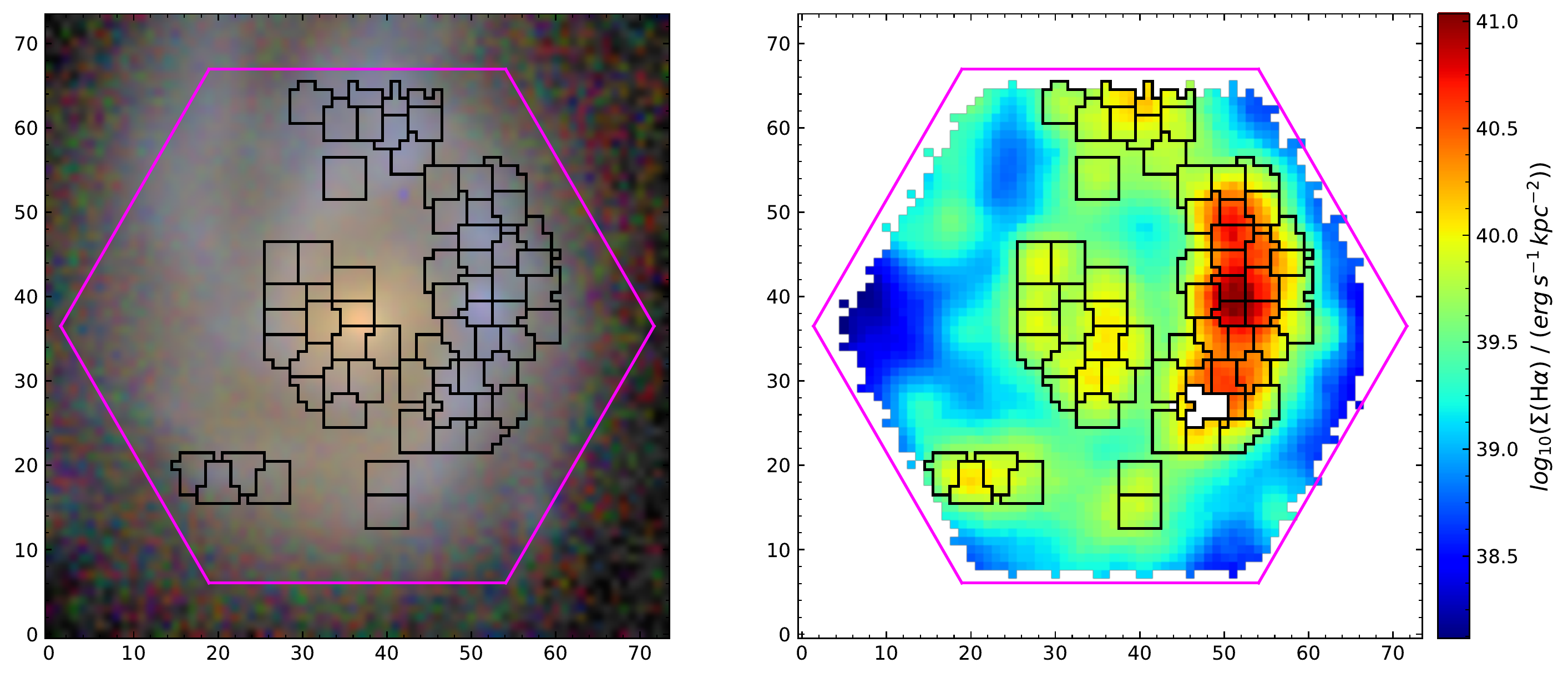}
    \caption{An example of {\hiiexplorer} applied to a MaNGA galaxy. Left panel: optical image of the galaxy with magenta hexagonal field of view of MaNGA. Right panel: {\ha} surface brightness map of this galaxy. Black boundaries of \ion{H}{2} regions are over-plotted in both panels.}
    \label{hii_example}
\end{figure}

In total, we have identified about 8000 \ion{H}{2} regions distributed
in 1155 galaxies, which is $\sim$25\% of the MPL-7 galaxy sample. In
what follows, these galaxies will be called ``star-forming galaxies''
and their \ion{H}{2} regions will form the parent catalog from which
the WR regions are identified. \autoref{hii_example} displays the
optical image and the $\Sigma H\alpha$ map with boundaries of \ion{H}{2} regions over-plotted for
one of the star-forming galaxies, as an example.  The fraction of
star-forming galaxies in our work is smaller than those from previous
studies.  For instance, \cite{Hsieh-SF} classified about half
of MaNGA MPL4 galaxies as star-forming galaxies, adopting a
specific star formation rate threshold. Therefore we would like to emphasize that, the fractions
of star-forming galaxies and \ion{H}{2} regions, as well as those of
WR galaxies/regions to be identified below, should be taken as
\textit{lower limits} of the real fractions.

\subsection{Spectrum stacking and full spectrum fitting}
\label{sec:fitting}

We stack the spectra within each \ion{H}{2} region to obtain an
average spectrum with high S/N. After applying {\drp} spectral masks, we
expect very few abnormal values in spectra.  Thus, we choose the
``weighted mean'' estimator for stacking, which is the statistically
optimal choice to lower the noise.  Spectra are weighted by their {\drp}
spectral error provided at each wavelength point.  We correct all
spectra to the rest-frame, considering both the
galaxy redshift and relative velocities of each spaxel. We choose
stellar velocity from {\dap} rather than gas velocity traced by emission
lines.  This is meant for a better alignment of stellar components
in \ion{H}{2} regions, in order for a better continuum fitting in
the next step.  
We understand this choice broadens the nebular emission lines in our stacked spectra, 
but the effects are minimal and insignificant for the scope of this work.
This process effectively reduces the noise of our
spectra, typically by about 25\%. Covariance is treated following the
formula in \citet[Figure 16]{Law-16} for derivation of the stacked
error. Basically, we first calculate the error without covariance by
the standard formula in weighted mean statistics (when assuming equal
input error, being a division of that error by
$\sqrt{N_{spectra}}$). Then we consider the effect of covariance by
multiplying that error by a factor from the formula in \citet{Law-16}.
The factor was obtained from a synthetic approach following \citet{2013A&A...549A..87H}. A mock observation of unity flux and Gaussian error was assumed and then put through data reduction, datacube resampling and spectral stacking. The standard deviation of the stacked spectrum is regarded as the real error and compared to the nominal error from error propagation without covariance, so as to obtain the correction factor. 

We then perform full spectral fitting to the stacked spectrum of each
\ion{H}{2} region. Our spectral fitting code is developed from
\citet{Li-05}. The basic idea was to utilize a set of eigenspectra as fitting
templates constructed by successive application of principal component
analysis (PCA). PCA was first applied to the observed stellar library STELIB \citep{STELIB} and some supplementary libraries. Then eigenspectra resulted from the first PCA run were used to fit a representive sample of galactic spectra selected from SDSS DR1, constituting fitting models for galactic spectra. Next, PCA is applied to the fitting models. Nine top eigenspectra from the second PCA run were obtained as the fitting templates to carry out subsequent full spectral fitting.
In this work, we re-construct the fitting templates by applying PCA to
the {\miles} single stellar population (SSP) models \citep{Vazdekis-10, Vazdekis-15}, which
includes 350 SSPs over 50 ages and 7
metallicity bins, assuming a Chabrier IMF \cite{Chabrier-03}.  We
adopt the first nine eigenspectra from the PCA as our fitting
templates.
For each \ion{H}{2} region, we perform full spectral fitting using
this code, iteratively masking out all significant emission lines as well as the
wavelength range of the WR blue bump ($4600-4750$ {\AA}). Details
about the fitting procedure and the emission line masking scheme
can be found in \citet{Li-05}.

\subsection{Identification of WR galaxies and WR regions}

For each \ion{H}{2} region we subtract the best-fit stellar spectrum
obtained from the previous subsection from the stacked spectrum.  We
then estimate the significance of the blue bump
{$\sigma_{\mbox{bump}}$}, defined as
\begin{equation}\label{eqn:sig_bump}
  \sigma_{\mbox{bump}} = \frac{\bar{f}_{\mbox{bump}} -
  \bar{f}_{\mbox{base}}}{f_{\mbox{rms}}},
\end{equation}
where $\bar{f}_{\mbox{bump}}$ is the average flux over the
  bump wavelength range $4600-4750$ {\AA}, $\bar{f}_{\mbox{base}}$
  is the average of the baseline, and $f_{\mbox{rms}}$ is
  the root mean square of the spectrum around the baseline
  over two wavelength windows beside the bump: $4492-4542$ {\AA}
  and $4760-4810$ {\AA}. The baseline should in principle be a
  horizontal line with zero flux thus giving rise to
  an average of $\bar{f}_{\mbox{base}}=0$, but the actual
  baseline always deviates from a zero line to some degrees
  due to imperfect spectral fitting. In order to take into account
  this effect, following \citet{Brinchmann-Kunth-Durret-08}, we determine
  a local baseline with a linear fitting of the residual spectrum
  over 4492-4542 {\AA} and 4760-4810 {\AA}, and use it when
  estimating $\sigma_{\mbox{bump}}$ with Eqn.~\ref{eqn:sig_bump}.

\begin{figure*}
	\includegraphics[width=1.0\textwidth]{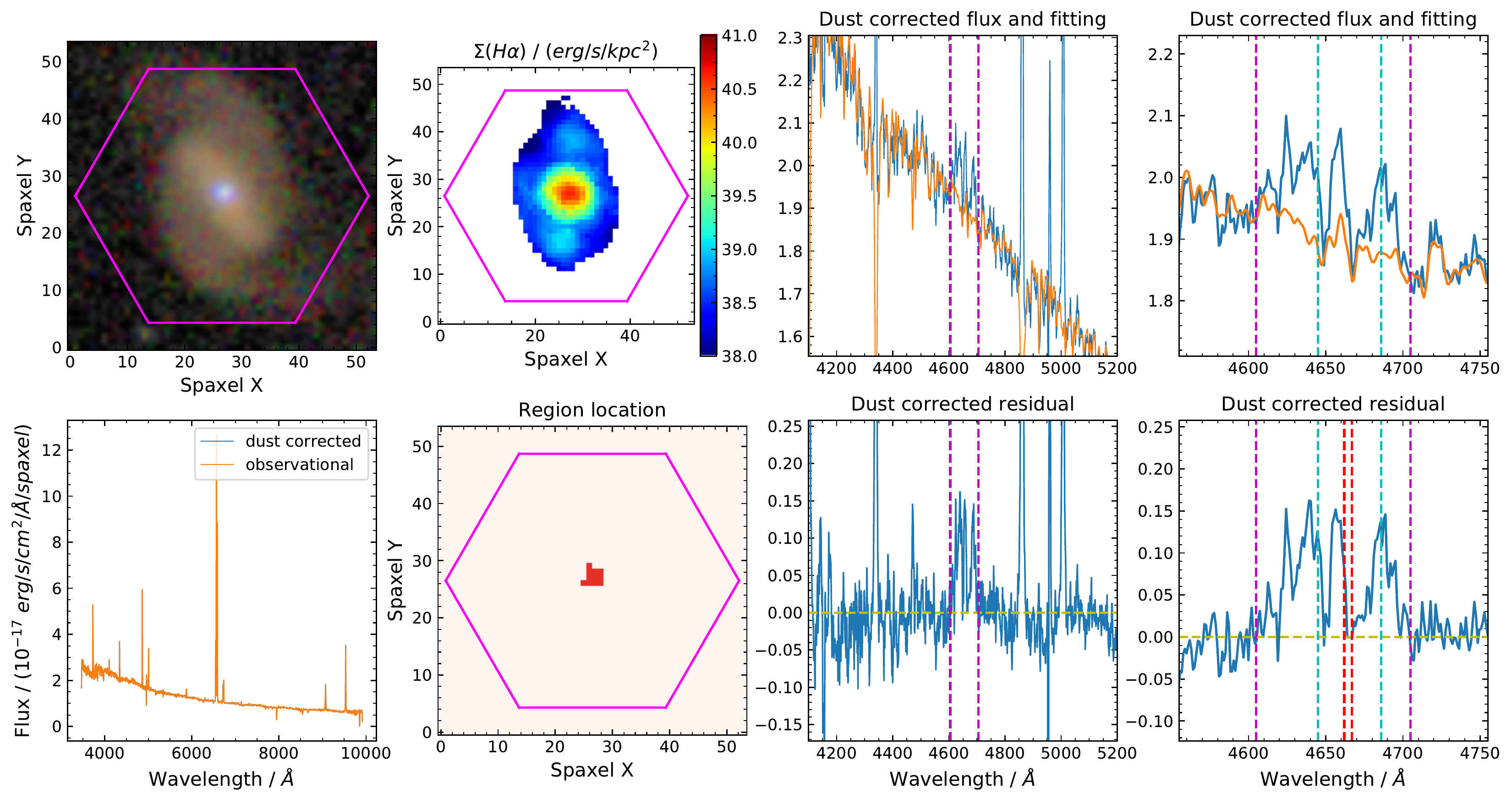}
    \caption{Example of visual inspection process. These combined panels are for determination of one \ion{H}{2} region. The eight panels involve the optical image (top left) and the {\ha} surface brightness map of the current galaxy, location (red area inside the hexagon in row 2 column 2), full original spectrum, zoomed-in original spectrum (one narrower and one wider) and zoomed-in residual (one narrower and one wider) of the current \ion{H}{2} region. Spectra have been smoothed with kernel=3.}
    \label{example}
\end{figure*}

  We consider all the \ion{H}{2} regions with $\sigma_{\mbox{bump}} >
  5$, ranking them by decreasing $\sigma_{\mbox{bump}}$.  We visually
  inspect the spectrum of each region, and classify them into four
  different catagories: a) a real WR region with broad emission
  components, b) a non-WR region with only narrow nebular emission
  lines, c) an active galactic nucleus (AGN) with obvious broad
  component in $H\alpha$ and/or $H\beta$ lines, or d) fluctuations due
  to noise. \autoref{example} shows an example galaxy which contains a
  WR region at the center. The figure includes the
  optical image and the {\ha} surface brightness map of the galaxy,
  location (red area inside the hexagon in row 2 column 2), full
  original spectrum, zoomed-in original spectrum (one narrower and one
  wider) and zoomed-in residual (one narrower and one wider) of the
  WR region. As can be seen, the spectrum of the WR region
  presents very strong nebular emission lines and relatively stronger emission
  at shorter wavelengths, indicative of strong ongoing star formation.

  We have visually examined about 3200 \ion{H}{2} regions, of which 267
  are identified to certainly present WR emission. The WR regions
  are distributed in 90 unique galaxies. We note that one of our
  WR galaxies is observed twice by MaNGA. In addition, out of the
  remaining \ion{H}{2} regions, we have further selected a subset
  of \textit{1609} \textit{tentative} WR regions whose spectra show possible WR profiles but are too weak for comfirmation at this point, 244 regions with only narrow 4686 {\AA} \ion{He}{2} emission and 57  suspicious quasi-WR
  features in red galaxies. These regions will be used for future studies.

  The final catalog of WR galaxies containing their basic properties
  is presented in \autoref{tab1}. The table's columns are arranged in the following order: (1) sequence number of WR galaxies in this catalog; (2) MaNGA-ID for unique idendification of MaNGA galaxies, (3) plate-ifu ID for MaNGA galaxies, (4) the number of WR regions contained in each WR galaxy, (5) right ascension, (6) declination, (7) redshift from NSA catalog, (8) stellar mass from K-correction fit for S{\'e}rsic fluxes from NSA catalog (we adopt h=0.7), (9) $T$-type morphological value from \citet{DomnguezSanchez-18}, (10) NUV-r color from NSA catalog.

\startlongtable
\begin{deluxetable*}{ccccccccccc}
%\centerwidetable
\centering
\tablecaption{WR galaxy catalog from MaNGA\label{WR table} \label{tab1}}
\tablewidth{700pt}
\tabletypesize{\scriptsize}
\tablehead{
\colhead{No.} & \colhead{MaNGA-ID} & \colhead{Plate-ifu} & \colhead{WR region } & \colhead{RA} &   \colhead{DEC} & \colhead{ Redshift} &  \colhead{Stellar mass} &  \colhead{T-type} &  \colhead{NUV-r} \\
\colhead{ } &  \colhead{} &  \colhead{}& \colhead{ number}&  \colhead{(Degree)}&   \colhead{(Degree)} & \colhead{} &  \dcolhead{(log_{10}(M/M_\odot))} & \colhead{} & \colhead{}
}
\startdata
   1 &  12-193481 &  7443-12703 &             5 &  229.52558 &  42.74584 &    0.0403 &           10.81 &         5.5 &   2.70 \\
   2 &  12-192116 &   7495-6102 &             2 &  204.51286 &  26.33821 &    0.0261 &            9.05 &         5.8 &   1.14 \\
   3 &    1-24357 &   7990-3703 &             3 &  262.09935 &  57.54541 &    0.0285 &            9.92 &         3.4 &   1.13 \\
   4 &    1-37084 &   8078-6104 &             6 &   42.73943 &   0.36941 &    0.0442 &            9.64 &         2.9 &   1.19 \\
   5 &   1-583411 &  8083-12702 &             1 &   50.24541 &  -0.36768 &    0.0210 &           11.09 &         5.1 &   2.95 \\
   6 &   1-604039 &  8083-12705 &             1 &   50.17898 &  -1.10865 &    0.0209 &           10.40 &         5.4 &   2.16 \\
   7 &   1-604748 &   8131-9101 &             3 &  112.57356 &  39.94208 &    0.0500 &           10.09 &         5.3 &   0.88 \\
   8 &   1-377321 &   8132-3702 &             4 &  110.55615 &  42.18364 &    0.0444 &            9.74 &         4.1 &   2.23 \\
   9 &    1-43505 &   8135-3704 &             2 &  114.89737 &  37.75151 &    0.0305 &            9.82 &         4.7 &   1.78 \\
  10 &    1-71137 &  8139-12702 &             4 &  113.50057 &  32.21576 &    0.0269 &            9.68 &         2.0 &   1.13 \\
  11 &   1-389720 &   8150-3701 &             1 &  147.58897 &  31.48794 &    0.0017 &            7.88 &         6.2 &   0.73 \\
  12 &   1-585731 &   8150-6103 &             1 &  147.14825 &  33.42162 &    0.0049 &           10.06 &         4.6 &   2.39 \\
  13 &    1-38819 &   8156-3701 &             2 &   55.59230 &  -0.58320 &    0.0524 &           10.13 &         5.0 &   1.48 \\
  14 &    1-52677 &   8158-1901 &             1 &   60.85933 &  -5.49184 &    0.0384 &            9.45 &         3.2 &   2.40 \\
  15 &   1-460288 &   8241-6102 &             3 &  126.05963 &  17.33195 &    0.0373 &           10.65 &         4.0 &   2.43 \\
  16 &    1-46577 &   8243-9101 &             4 &  128.17838 &  52.41678 &    0.0433 &            9.89 &         4.2 &   1.74 \\
  17 &   1-137961 &   8249-3703 &             1 &  139.72047 &  45.72778 &    0.0264 &           10.02 &        -1.3 &   3.52 \\
  18 &   1-137875 &   8249-6102 &             3 &  137.33592 &  45.06551 &    0.0510 &           10.32 &         3.7 &   1.90 \\
  19 &   1-217300 &   8250-3703 &             5 &  139.73996 &  43.50058 &    0.0401 &            9.52 &        -1.6 &   1.54 \\
  20 &   1-217221 &   8250-6101 &             2 &  138.75315 &  42.02439 &    0.0279 &           10.41 &         4.3 &   2.63 \\
  21 &   1-585641 &   8252-6103 &             2 &  144.15499 &  48.47438 &    0.0259 &           10.57 &         4.0 &   3.15 \\
  22 &   1-138157 &   8252-9102 &             3 &  145.54153 &  48.01286 &    0.0562 &           10.19 &         3.9 &   1.79 \\
  23 &   1-255959 &   8256-9102\tablenotemark{a} &             7 &  165.10414 &  43.01969 &    0.0375 &            9.89 &         7.0 &   0.75 \\
24 &   1-277290 &  8257-12701 &             4 &  165.49582 &  45.22802 &    0.0200 &           10.65 &         5.4 &   2.32 \\
  25 &   1-277293 &   8257-3704 &             1 &  165.55361 &  45.30387 &    0.0202 &            9.00 &        10.0 &   3.86 \\
  26 &   1-256496 &   8258-3704 &             1 &  167.02504 &  43.89461 &    0.0585 &           10.26 &         3.8 &   1.73 \\
  27 &   1-282147 &  8261-12703 &             2 &  184.35779 &  46.56687 &    0.0235 &            9.55 &         5.9 &   1.41 \\
  28 &   1-258306 &   8262-3701 &             1 &  183.57898 &  43.53528 &    0.0241 &            9.77 &         3.9 &   2.23 \\
  29 &   1-589908 &   8262-9102 &             1 &  184.55357 &  44.17324 &    0.0245 &           10.44 &         5.2 &   1.99 \\
  30 &   1-628628 &   8309-3703 &             6 &  210.62359 &  54.27100 &    0.0005 &            8.13 &       -99.0 &   0.39 \\
  31 &   1-248388 &  8313-12702 &             3 &  240.67742 &  41.19726 &    0.0333 &           10.59 &         4.8 &   2.30 \\
  32 &   1-248352 &   8313-1901 &             2 &  240.28713 &  41.88075 &    0.0243 &            9.19 &         1.9 &   1.59 \\
  33 &   1-523050 &   8320-9101 &             6 &  206.31385 &  23.31651 &    0.0297 &           10.22 &         4.7 &   1.95 \\
  34 &   1-419028 &   8322-3701 &             3 &  199.06648 &  30.26453 &    0.0492 &           10.95 &         4.6 &   2.30 \\
  35 &   1-591611 &   8322-9101 &             6 &  199.60756 &  31.46795 &    0.0187 &            9.84 &         5.3 &   1.58 \\
  36 &   1-591379 &  8323-12701 &             1 &  196.37038 &  33.84872 &    0.0238 &            9.84 &         6.7 &   2.20 \\
  37 &   1-234997 &  8325-12702 &             1 &  209.89514 &  47.14768 &    0.0420 &            9.59 &         6.3 &   1.55 \\
  38 &   1-266045 &   8329-3702 &             1 &  213.49543 &  43.89295 &    0.0403 &            9.72 &         3.6 &   2.50 \\
  39 &   1-491225 &   8338-6102 &             2 &  172.68267 &  22.36354 &    0.0224 &            9.56 &         5.0 &   2.38 \\
  40 &   1-575796 &  8341-12705 &             2 &  191.49328 &  45.19901 &    0.0251 &            9.94 &         5.8 &   1.75 \\
  41 &   1-156037 &   8439-9102 &             2 &  143.75402 &  48.97674 &    0.0250 &            9.33 &         5.3 &   1.43 \\
  42 &   1-591580 &   8442-3701 &             1 &  199.21314 &  31.58114 &    0.0298 &           10.46 &         4.9 &   2.84 \\
  43 &   1-419153 &   8442-6102 &             1 &  199.21116 &  31.63067 &    0.0303 &            9.89 &         6.2 &   2.01 \\
  44 &   1-418242 &   8446-3703 &             1 &  205.58417 &  36.95363 &    0.0215 &            9.77 &         4.2 &   2.03 \\
  45 &   1-488712 &   8449-3703 &             4 &  169.29926 &  23.58566 &    0.0421 &            9.50 &         4.8 &   1.77 \\
  46 &   1-489814 &   8450-3703 &             2 &  170.44434 &  22.48082 &    0.0351 &           10.37 &         3.7 &   2.04 \\
  47 &   1-608252 &  8454-12703 &            15 &  154.77137 &  46.45411 &    0.0307 &           10.74 &         4.6 &   2.18 \\
  48 &   1-254342 &  8455-12701 &             4 &  154.73181 &  40.61294 &    0.0292 &           10.13 &         5.3 &   2.29 \\
  49 &   1-275176 &   8455-9101 &             3 &  157.18363 &  39.77859 &    0.0300 &            9.97 &         5.5 &   1.95 \\
  50 &   1-585744 &   8458-3702 &             9 &  147.56250 &  45.95731 &    0.0249 &            9.75 &         3.9 &   0.99 \\
  51 &   1-284048 &   8465-3701 &            11 &  195.31932 &  48.06019 &    0.0300 &           10.22 &         3.5 &   1.68 \\
  52 &   1-234092 &   8465-6102 &             1 &  197.54970 &  48.62339 &    0.0283 &            9.60 &         3.1 &   1.53 \\
  53 &   1-209198 &   8485-3702 &             2 &  233.72546 &  47.76180 &    0.0230 &            9.65 &         4.4 &   1.78 \\
  54 &    1-93305 &   8549-6104 &             1 &  244.40158 &  46.08200 &    0.0196 &            9.57 &         2.1 &   1.97 \\
  55 &   1-247373 &   8551-1902 &             5 &  234.59171 &  45.80194 &    0.0214 &            8.74 &         1.7 &   1.91 \\
  56 &    1-91019 &   8553-3704 &             1 &  234.97035 &  56.36832 &    0.0459 &            9.88 &         4.2 &   1.92 \\
  57 &   1-584598 &   8566-3704 &             6 &  115.22481 &  40.06964 &    0.0416 &            9.76 &         4.3 &   1.01 \\
  58 &   1-274368 &   8568-3703 &             1 &  155.69307 &  37.67347 &    0.0226 &            9.38 &        -0.6 &   1.90 \\
  59 &   1-634138 &  8588-12702 &             1 &  250.31305 &  39.29009 &    0.0305 &           10.18 &         4.4 &   2.10 \\
  60 &   1-136286 &   8606-9102 &             5 &  255.70905 &  36.70675 &    0.0328 &           10.05 &         5.1 &   2.08 \\
  61 &   1-177270 &  8613-12703 &             3 &  256.81775 &  34.82261 &    0.0367 &            9.95 &         3.3 &   1.62 \\
  62 &   1-178686 &  8623-12703 &             1 &  309.98326 &   0.97615 &    0.0522 &           10.43 &         0.1 &   3.63 \\
  63 &    1-24423 &  8626-12704 &             6 &  263.75522 &  57.05243 &    0.0472 &            8.88 &         3.5 &   0.14 \\
  64 &   1-379291 &   8712-6101 &             1 &  118.72692 &  53.84628 &    0.0349 &           10.84 &         3.5 &   2.68 \\
  65 &   1-379410 &   8712-6103 &             1 &  120.23007 &  53.67056 &    0.0406 &            9.86 &         2.5 &   2.10 \\
  66 &    1-71974 &   8713-9102 &             7 &  118.85539 &  39.18609 &    0.0332 &           10.66 &         4.8 &   1.33 \\
  67 &   1-121735 &   8717-3703 &             1 &  118.31817 &  35.57258 &    0.0458 &            9.98 &         1.3 &   2.25 \\
  68 &    1-44745 &  8719-12702 &             3 &  120.19928 &  46.69053 &    0.0194 &            9.80 &         5.6 &   2.03 \\
  69 &    1-51766 &   8727-3702 &             4 &   54.55405 &  -5.54040 &    0.0221 &            8.79 &        -1.2 &   1.48 \\
  70 &   1-456306 &   8932-3701 &             2 &  194.65534 &  27.17656 &    0.0256 &            9.73 &         4.1 &   1.87 \\
  71 &   1-456772 &   8934-3701 &             6 &  194.02549 &  27.67798 &    0.0165 &            9.23 &        10.0 &   1.58 \\
  72 &   1-298533 &   8939-6102 &             3 &  124.84327 &  23.74728 &    0.0153 &            9.36 &         5.4 &   1.97 \\
  73 &   1-164148 &   8941-3702 &             3 &  120.00802 &  27.11454 &    0.0426 &            9.75 &         4.2 &   1.84 \\
  74 &   1-392670 &  8943-12704 &             1 &  156.43286 &  36.02359 &    0.0538 &           10.11 &         3.5 &   1.57 \\
  75 &   1-279617 &   8945-3702 &             5 &  173.36190 &  47.28673 &    0.0456 &            9.85 &        -0.9 &   1.59 \\
  76 &   1-153038 &   8977-3704 &             2 &  118.77440 &  32.72867 &    0.0178 &            9.02 &        -0.3 &   1.67 \\
  77 &   1-457811 &   8982-9101 &             4 &  201.71365 &  26.59124 &    0.0235 &           10.09 &        10.0 &   2.29 \\
  78 &   1-386685 &   8987-9101 &             1 &  137.44943 &  27.86231 &    0.0204 &            9.25 &         4.9 &   2.00 \\
  79 &   1-174914 &   8990-6104 &             1 &  174.74526 &  50.00589 &    0.0466 &           10.08 &         4.5 &   1.79 \\
  80 &   1-314332 &   9024-1902 &             2 &  224.43779 &  33.16563 &    0.0300 &            9.70 &         2.6 &   1.65 \\
  81 &    1-94584 &   9026-3701 &             1 &  250.16267 &  43.34609 &    0.0228 &            9.60 &        -2.3 &   3.08 \\
  82 &   1-199432 &   9037-9101 &             1 &  234.62042 &  43.73320 &    0.0184 &            9.31 &         4.0 &   2.15 \\
  83 &   1-153938 &  9183-12705 &             2 &  123.10611 &  37.73022 &    0.0385 &           10.42 &        -1.9 &   2.30 \\
  84 &    1-37863 &  9193-12704 &             7 &   46.66491 &   0.06198 &    0.1074 &           10.76 &        -0.9 &   2.14 \\
  85 &    1-45151 &  9487-12702 &             1 &  122.79107 &  45.66357 &    0.0229 &            9.74 &         5.1 &   2.23 \\
  86 &   1-382712 &   9491-6101 &             1 &  119.17438 &  17.99117 &    0.0412 &           10.79 &         4.3 &   2.16 \\
  87 &   1-386150 &   9506-6102 &             1 &  133.75984 &  26.67535 &    0.0274 &           10.04 &         4.6 &   2.26 \\
  88 &   1-298835 &   9508-1901 &             2 &  126.08120 &  25.67447 &    0.0282 &            9.85 &        -0.2 &   1.72 \\
  89 &   1-218233 &   9509-3702 &             3 &  122.43975 &  25.88031 &    0.0251 &            9.58 &         4.7 &   1.27 \\
  90 &   1-594855 &   9883-3701 &             6 &  255.13405 &  32.67077 &    0.0325 &           10.37 &         4.3 &   2.30 \\
\enddata
\tablenotetext{a}{This galaxy has repeat observation. The other plate-ifu ID is 8274-9102.}
\end{deluxetable*}

\section{Global properites of WR host galaxies}
\label{sec:global}

\subsection{Stellar mass function of WR galaxies}

\begin{figure}
	\includegraphics[width=\columnwidth]{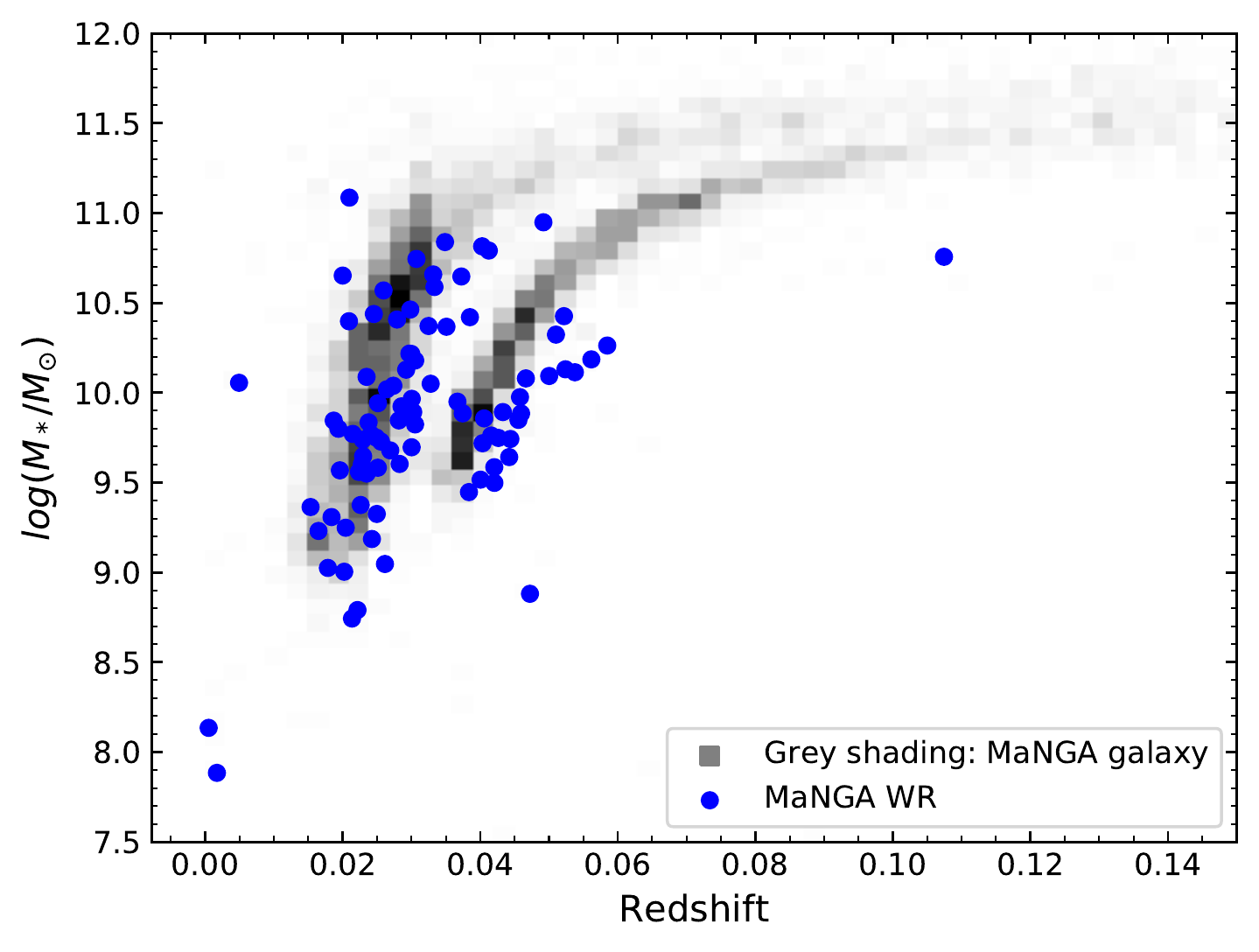}
    \caption{Distribution of WR galaxies in redshift-mass diagram. The grey shading represents the distribution of MaNGA galaxies. The blue dots are WR galaxies in our catalog.}
    \label{mass-z}
\end{figure}

\autoref{mass-z} displays the WR galaxies in the plane of stellar mass (assuming h=0.7)
versus redshift. For comparison, the distribution of all galaxies in
MaNGA MPL-7 is plotted as grey-scale background.
Stellar masses and redshifts of all galaxies 
in this figure are taken from the the NSA
(see \S~\ref{sec:data_and_selection}). MaNGA MPL-7 galaxies are
distributed in two narrow bands which correspond to the Primary Sample
(lower redshifts at given mass) and the Secondary Sample (higher
redshifts at given mass) as described in \S~\ref{sec:manga_overview}.
Our WR catalog generally follows the distribution of parent MaNGA
sample, but biased to lower redshifts with $z\lesssim0.06$ and
intermediate-to-low masses with $M_\ast\lesssim10^{11}M_\odot$. This
might be reflecting the fact that WR regions of similar sizes can be
more visible if hosted by more nearby galaxies. Furthermore,
at fixed redshift, the WR galaxies on average appear to be less
massive than MaNGA MPL-7 galaxies, an effect that is more
pronounced for the Secondary Sample. This might be attributed to
selection bias, or a real trend for WR regions to be preferentially
found in relatively low-mass galaxies. We will come back to this
point in a later subsection.

\begin{figure}
	\includegraphics[width=\columnwidth]{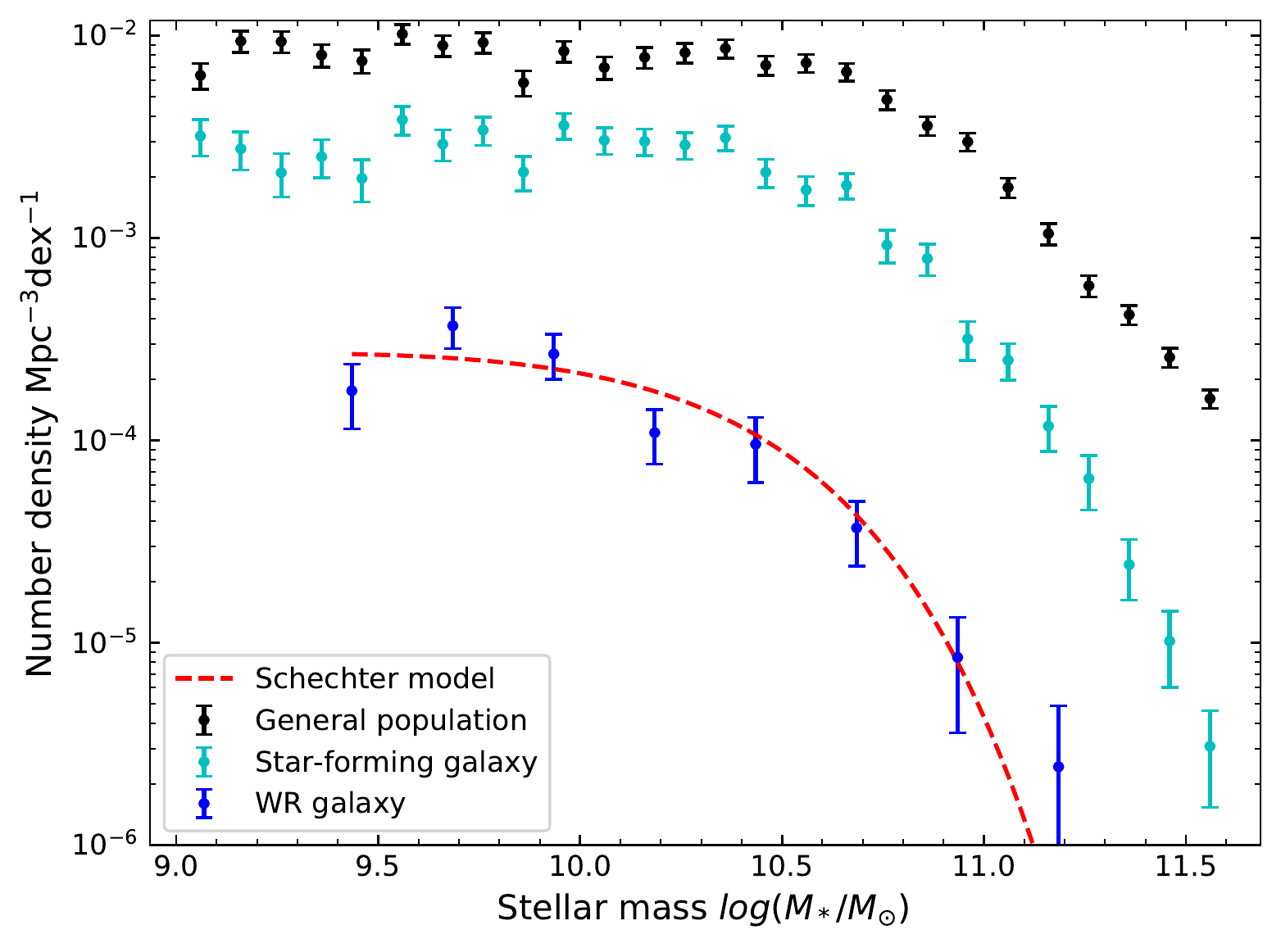}
    \caption{The stellar mass function of the WR catalog (blue dots)
      in comparison with the general galaxy population constructed from MaNGA Primary Sample (black dots) and star-forming galaxies defind in this work (see \S~\ref{hii_region}; cyan dots). All samples have been applied the volume correction provided in
      \citet{Wake-17}.}
    \label{mass-function}
\end{figure}

Taking advantage of the relatively large sample size  as well as the
well-understood selection effects of our WR catalog, we have estimated
for the first time the stellar mass function of WR galaxies, plotted
as blue symbols in \autoref{mass-function}. For comparison, we have
estimated the stellar mass function of the general population of
galaxies using MaNGA MPL-7 sample, and that of star-forming
galaxies using our \ion{H}{2} catalog. When estimating the stellar
mass functions, we have corrected the effect of sample incompletness
due to selections using the weights accompanying the MPL-7 release and
described in detail in the Appendix of \cite{Wake-17}. We only use the Primary Sample and its corresponding weights in \autoref{mass-function} for better statistics at the low mass end.
The errors of
the mass functions are Poisson counting error.
In \autoref{WR mass function} we tabulate the stellar mass
function estimate of our WR galaxy catalog and its ratio to the
general galaxy population. Like the stellar mass function of general
population, the mass function of WR galaxies can be well described by
a Schechter function \citep{1976ApJ...203..297S}.  In the figure we
plot the best-fit Schechter function, for which the three parameters
are: amplitude $\phi_*$ = 0.000157 Mpc$^{-3}$, characteristic mass
$log_{10}(M^*/M_\odot)$ = 10.332, and the faint-end slope $\alpha$ =
-0.905. The integral of this Schechter function
over the mass range of $10^9M_\odot<M_\ast<10^{11.5}M_\odot$ 
gives an average number density of $3.47\times 10^{-4}Mpc^{-3}$ for the WR galaxies in
the Local Universe. This should be regarded as a lower limit of the
real number density, considering that we may have missed some
weak WR regions due to the limited data quality and selection effects.

\begin{deluxetable}{cccccc}
\tablecaption{Stellar mass function of MaNGA WR galaxies \label{WR mass function}}
\tabletypesize{\scriptsize}
\tablehead{
\colhead{No.} & \colhead{Mass} & \colhead{$\phi$} & \colhead{$\sigma(\phi)$} & \colhead{WR fraction} &   \colhead{$\sigma$(fraction)}\\
\colhead{ } &  \colhead{$log_{10}(M_*/M_\odot)$} &  \colhead{Mpc$^{-3}$dex$^{-1}$}& \colhead{ }&\colhead{ }&\colhead{ }
}
\startdata
1 & 9.18-9.68 & 1.73E-04 & 6.12E-05 & 1.87\% & 0.66\% \\
2 & 9.43-9.93 & 3.62E-04 & 8.31E-05 & 3.60\% & 0.83\% \\
3 & 9.68-10.18 & 2.63E-04 & 6.58E-05 & 3.12\% & 0.78\% \\
4 & 9.93-10.43 & 1.08E-04 & 3.24E-05 & 1.48\% & 0.45\% \\
5 & 10.18-10.68 & 9.43E-05 & 3.33E-05 & 1.37\% & 0.49\% \\
6 & 10.43-10.93 & 3.64E-05 & 1.29E-05 & 0.71\% & 0.25\% \\
7 & 10.68-11.18 & 8.34E-06 & 4.81E-06 & 0.26\% & 0.15\% \\
8 & 10.93-11.43 & 2.40E-06 & 2.40E-06 & 0.18\% & 0.18\% \\
\enddata
\end{deluxetable}

\begin{figure}
	\includegraphics[width=\columnwidth]{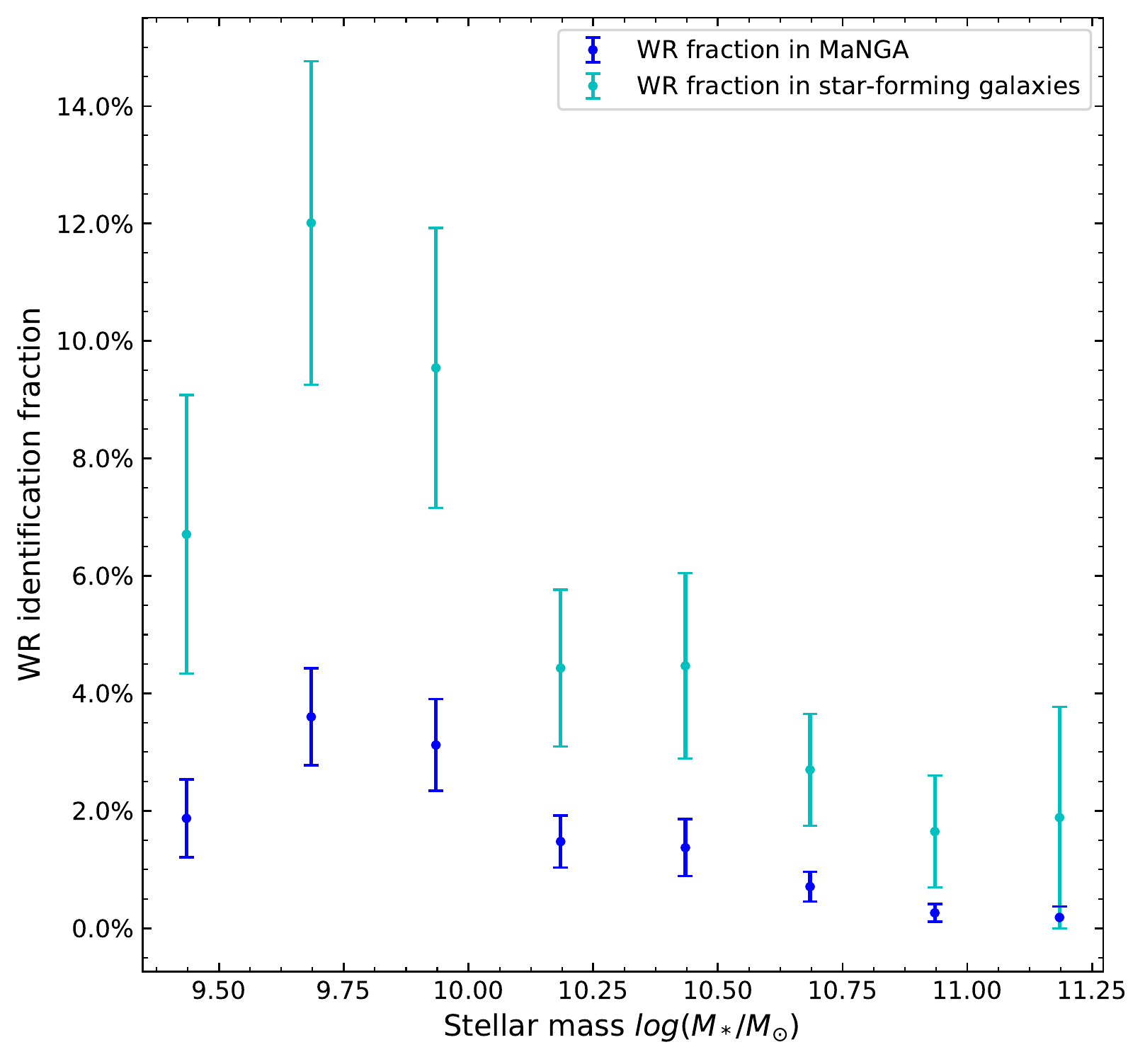}
    \caption{The detection fraction of WR galaxies in different mass bins, with respect to the general galaxy population (blue symbols) and the star-forming galaxy sample defined in this work (see \S~\ref{hii_region}; cyan symbols). The errorbar shows Poisson counting error.}
    \label{fraction-mass}
\end{figure}

\begin{figure}
	\includegraphics[width=\columnwidth]{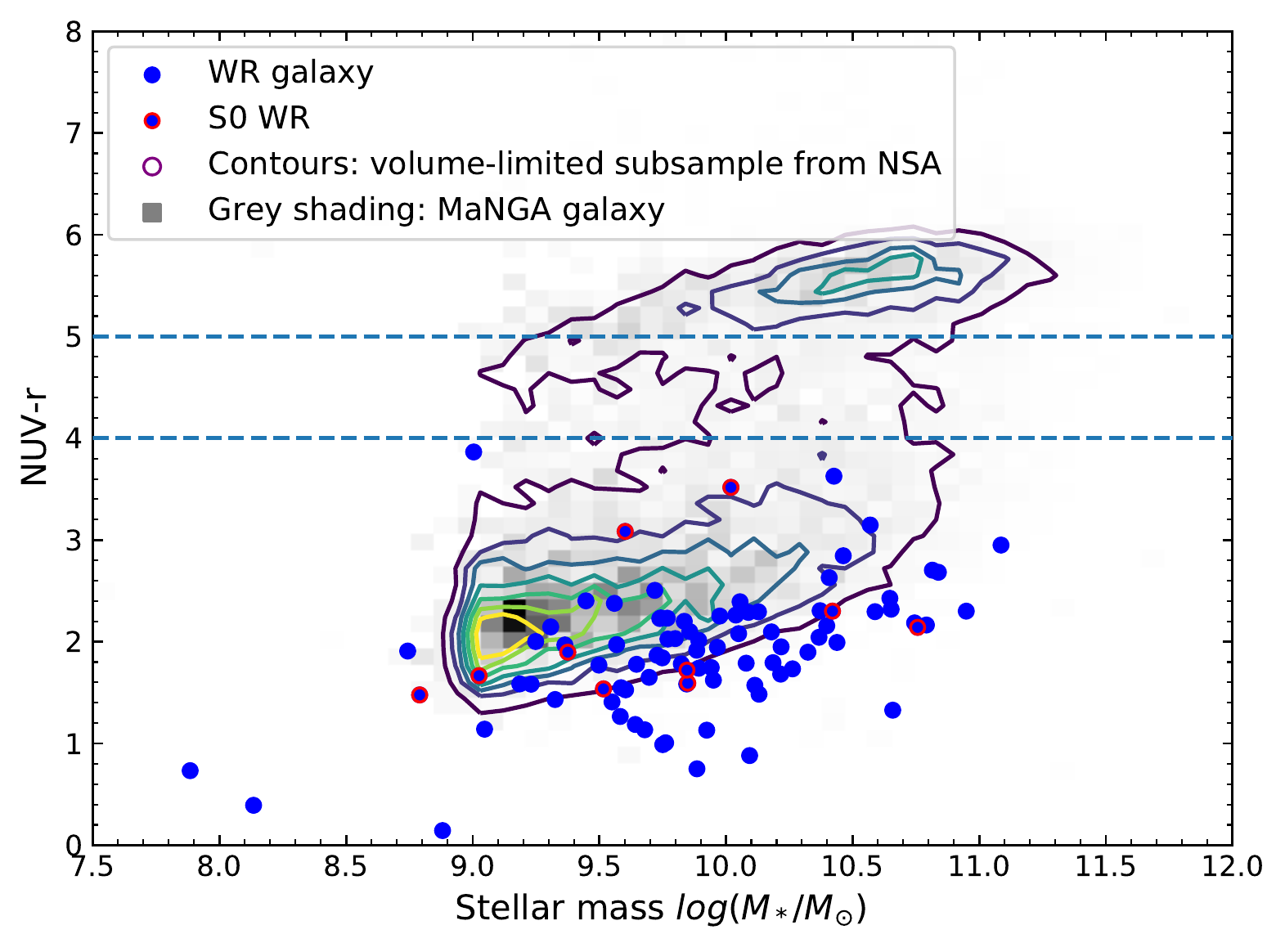}
    \caption{The NUV-r colour-mass diagram. The contours are a volume limited subsample from NSA catalog ({$z<0.03$} and $log_{10}(M_*/M_\odot)>9$). The grey shading is MaNGA sample corrected by galaxy weights. The blue dots are WR galaxies with S0 type WR galaxies highlighted with an additional red circle. The color and mass values are from NSA catalog. The cyan dotted lines indicate the empirical boundaries of blue cloud galaxies (lower), green valley (middle) and red sequence galaxies (upper).}
    \label{color-mass}
\end{figure}

As can be seen from \autoref{mass-function}, the stellar mass function
of WR galaxies is well determined over about two orders of magnitude
in mass, from $\sim10^{9}M_\odot$ up to $\sim10^{11}M_\odot$. The WR
galaxy sample appears to keep the same shape over the entire mass
range as the general population in terms of both the flat slope at the
low-mass end and the sharp decline at the massive end. However, as
expected, the amplitude of the mass function of WR galaxies is much
lower than the whole galaxy population, with a ratio of $\sim10^{-2}$
at all masses. Comparing the number density of WR galaxies as given by
the Schechter function above with the number density from the stellar
mass function of general galaxies \citep[e.g.][]{Li-White-09}, we
estimate that the average WR galaxy fration is 1.4\%.  This
result echos the known and expected fact that WR galaxies form a
rather rare population in the Local Universe.  This is more clearly
shown in \autoref{fraction-mass} which plots the ratio of the mass
function of the WR sample with respect to the general galaxy
population (blue symbols) and the star-forming galaxy sample (cyan
symbols). The WR population is most abundant at
$M_\ast\sim10^{9.7}M_\odot$, with a maximum fraction of $\sim$4\%. The
fraction of WR galaxies decreases at both higher and lower masses,
down to $\sim2$\% at $\sim10^{9.5}M_\odot$ and $\sim0.2$\% at above
$10^{11}M_\odot$.

\subsection{Stellar population properties}

\autoref{color-mass} displays the WR galaxies on the $NUV-r$ versus
stellar mass diagram. MaNGA MPL-7 sample is plotted as grey-scale
background for comparison.  We have corrected the effect of MaNGA
sample selection by weighting the MPL-7 galaxies using the weights
provided by \citet{Wake-17}. In addition, we
have selected a volume-limited sample of galaxies from the NSA with
redshifts $z<0.03$ and stellar masses $M_\ast>10^{9}M_\odot$. The
distribution of this sample is shown as contours in the figure.  The
well-known bimodal distribution of galaxies in the color space is
clearly seen in both the MaNGA and the NSA sample, where the galaxies
are separated into two populations: the red sequence with $NUV-r\ga 5$
and the blue cloud with $NUV-r\la 4$, with an intermediate population
falling in the green valley.  In contrast, the WR galaxies are found
exclusievely in the blue cloud, mostly with $NUV-r\la 3$. The WR
galaxies are more massive than the NSA galaxies of similar colors.

\begin{figure}
	\includegraphics[width=\columnwidth]{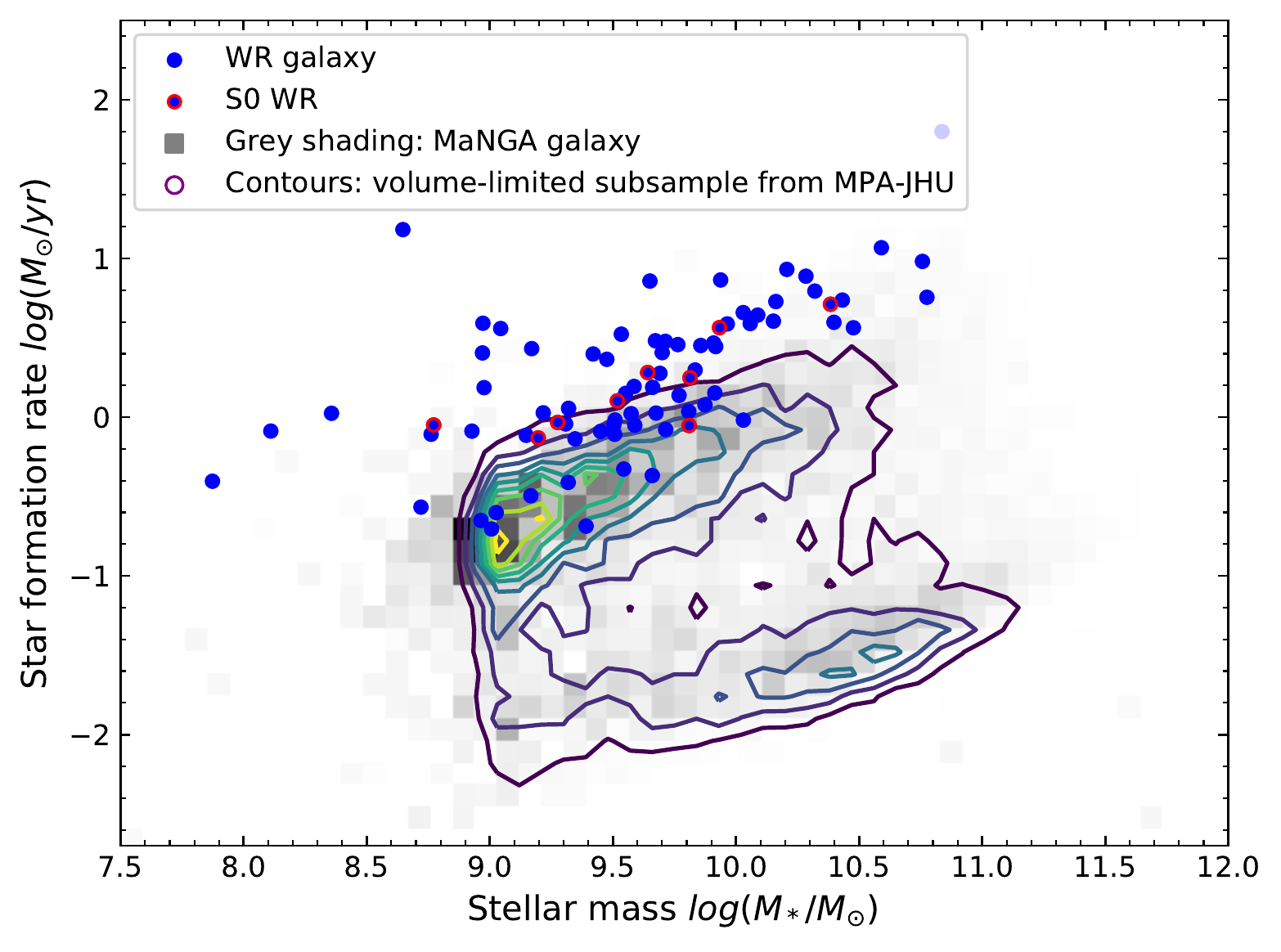}
    \caption{Star formation rate versus stellar mass. The contours are a volume limited subsample drawn from MPA-JHU catalog. The grey shading is MaNGA sample corrected by weights and the blue dots are WR galaxies with S0 type WR galaxies highlighted with an additional red circle. All values for this figure are from MPA-JHU catalog.}
    \label{sfr_mass}
\end{figure}

\autoref{sfr_mass} displays the distribution of the same samples
in the diagram of star formation rate (SFR) versus stellar mass.
Estimates of SFRs are taken from the MPA-JHU SDSS database
\footnote{\url{https://wwwmpa.mpa-garching.mpg.de/SDSS/DR7/}},
provided by \citet{Brinchmann-04}. The same weighting correction is applied to MaNGA galaxies and the same criteria as previous is adopted for a volume-limited subsample drawn from MPA-JHU catalog. Similarly to the previous figure,
the bimodality of general galaxies from both MaNGA MPL-7 and MPA-JHU
is well reproduced, with galaxies of higher SFRs at fixed mass
falling in the star-forming main sequence and those of lower SFRs
falling in the quenched sequence. WR galaxies are located at the
top end of the general-population, with highest SFRs at fixed mass.
The two figures are consistent with each other, telling us a simple,
expected fact that WR regions are found exclusively in strongly
star-forming galaxies which are predominantly blue. 

\subsection{Morphology and structural properties}

\begin{figure}
	\includegraphics[width=\columnwidth]{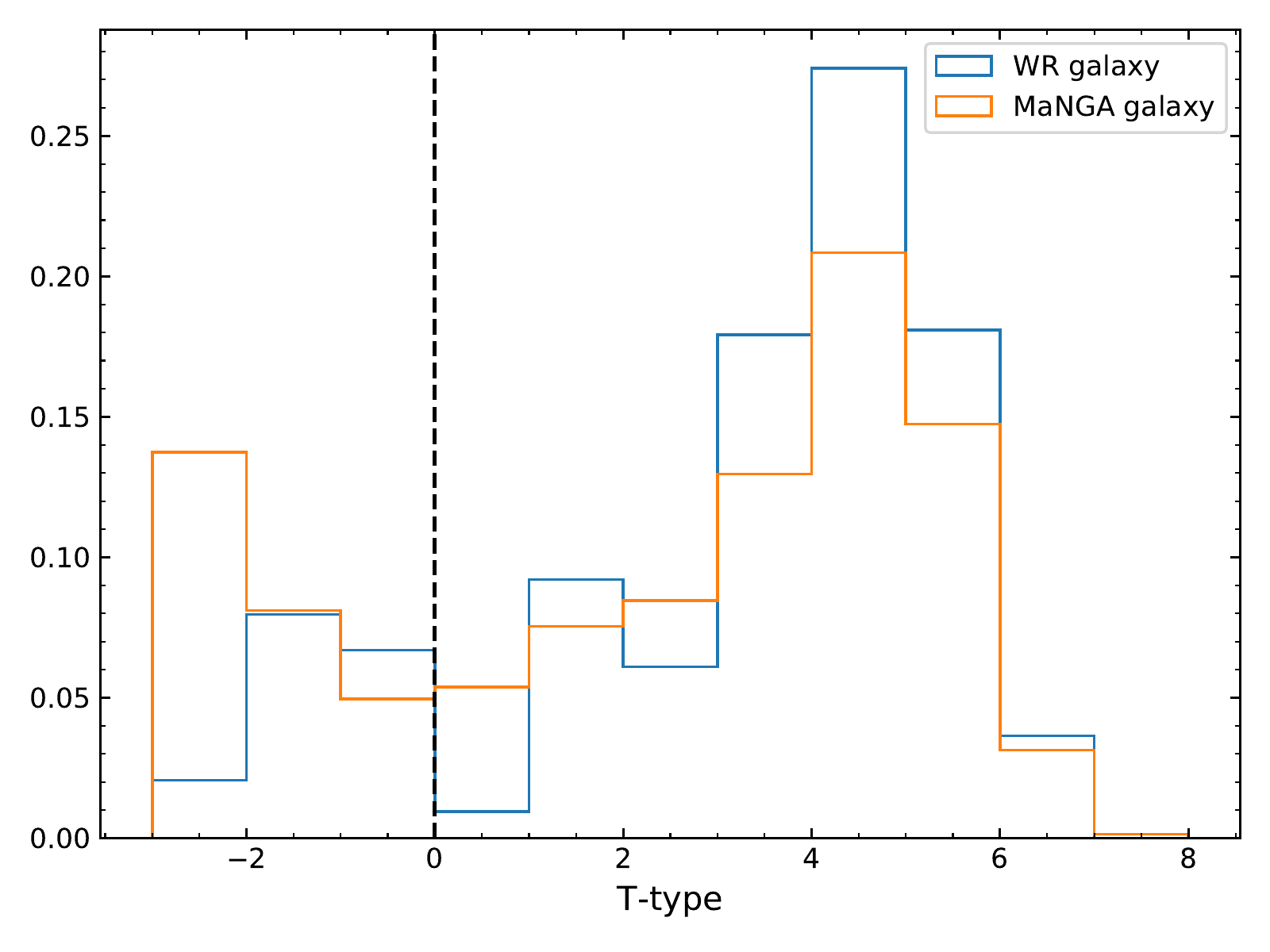}
    \caption{The morphology histogram of MaNGA and WR population. Both sample have been applied the volume correction. $T$-type value is an indicator of morphology type with negative values indicating early-type morphologies and positive values being late-type morphologies. Errorbars are Poisson counting error. All early-type WR galaxies are found to be S0 galaxies.}
    \label{morphology}
\end{figure}

Now we examine the morphology and structural properties of the WR
galaxies. For this, we adopt the morphology classification from
\citet{DomnguezSanchez-18} which determines the $T$-type value for
each galaxy in SDSS DR7 by Convolutional Neural Networks (CNNs). The
input of the CNNs are raw RGB cutouts from SDSS and they trained the
CNNs with two visual classification catalogues: GZ
\citep{2013MNRAS.435.2835W} and \cite{2010ApJS..186..427N}. Galaxies
with a negative $T$-type value usually have an early-type morphology
(elliptical or S0 type), while a positive $T$-type value indicates a
late-type galaxy. \autoref{morphology} shows the histogram of $T$-type
values  for both MaNGA MPL-7 sample and our WR galaxy catalog. We have
corrected the effect of sample selection for both samples using the
weights from \citet{Wake-17} as in previous figures.  We find both
samples to cover the full range of $T$-type, while the WR sample tend
to have a slightly higher fraction of late-type galaxies. This can be
understood considering that the majority of the WR sample are
star-forming galaxies (see above). Out of the 90 WR galaxies in our
sample, 10 are negative in $T$-type, indicative of early-type
morphology.  By further checking the S0 probability provided by
\citet{DomnguezSanchez-18} and visually examining their optical
images, we find all of them are lenticular (S0-type) galaxies. These
galaxies are highlighted with red circles in \autoref{color-mass} and
\autoref{sfr_mass}, as well as the following \autoref{gini-m20}.

%\begin{figure*}
%  \centering
%  \includegraphics[width=0.8\textwidth]{sersic-mass.pdf}
%  \caption{The distribution of S{\'e}rsic index n vs mass in the main panel with two side panels of S{\'e}rsic index histograms for two ranges of galaxy mass. In the main panel, the contours are a volume-limited subsample from NSA catalog (contour levels not linearly separated), the grey shading is MaNGA sample corrected by galaxy weights, and the blue dots are WR galaxies with S0 type WR galaxies highlighted with additional red circles. Two side panels show the histograms of S{\'e}rsic index for the aforementioned three samples, one panel for $M_*>10^{10}M_\odot$ for all three samples and the other for $M_*<10^{10}M_\odot$. Errorbars are Poisson counting error. NSA-based and MaNGA-corrected samples overlap at some bins, for they both reflect the general population.}
%    \label{sersic_mass2}
%\end{figure*}
% $n=6$ is the manual upper limit in fitting and this causes some pile-up at the top.

\begin{figure*}
\centering
\includegraphics[width=\textwidth]{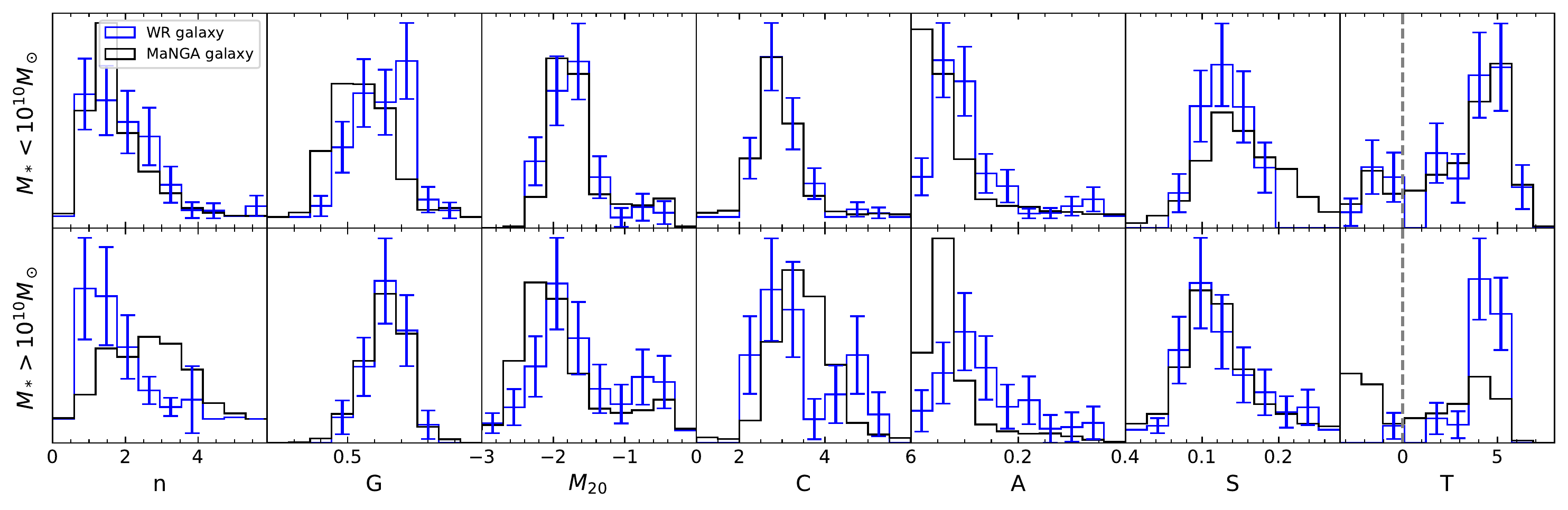}
\caption{Distribution of structural parameters of WR galaxies (in blue) in comparison with general population reconstructed from MaNGA galaxies with weights (in black). Errorbars are Poisson counting error. The upper panels are for low-mass galaxies with $M_*<10^{10}M_\odot$ while the lower panels are for high-mass galaxies with $M_*>10^{10}M_\odot$. Panels from left to right are for different structural parameters: S{\'e}rsic index $n$, Gini index $G$, second-order moment of light $M_{20}$, concentration index $C$, rotational asymmetry A, and clumpiness S, as well as the morphological $T$-type. The vertical dashed lines in $T$-type panels indicate the empirical division between late-type (positive $T$) and early-type (negative $T$) galaxies. All early-type WR galaxies are also classified as S0 lenticular galaxies.}
\label{structural}
\end{figure*}

\begin{figure}
\centering
\includegraphics[width=\columnwidth]{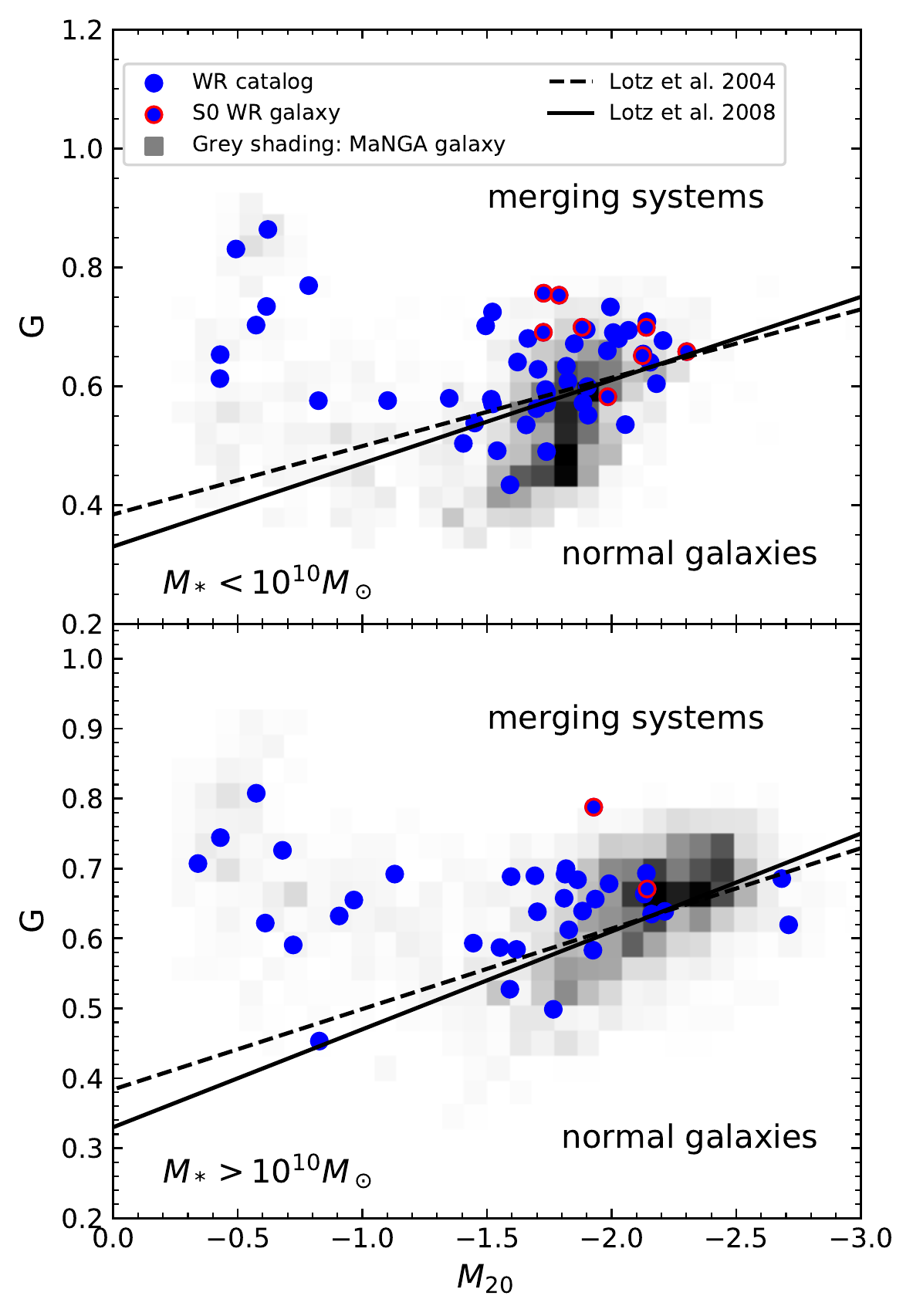}
\caption{The $G$ versus $M_{20}$ diagram for the WR galaxies with low
  (upper) and high (lower) stellar masses. The grey-scale background
  is the distribution of general population of galaxies reconstructed
  from the MaNGA MPL-7 sample with weights. Blue dots are WR galaxies
  with S0-type WR galaxies highlighted with an additional red
  circle. Two diagnostic lines separating merging systems from normal
  galaxies are from \citet[][solid]{2008ApJ...672..177L} and
  \citet[][dashed]{2004AJ....128..163L}.}
\label{gini-m20}
\end{figure}

%\begin{figure*}
%  \centering
%  \includegraphics[width=0.8\textwidth]{sersic-mass.pdf}
%  \caption{The distribution of S{\'e}rsic index n vs mass in the main panel with two side panels of S{\'e}rsic index histograms for two ranges of galaxy mass. In the main panel, the contours are a volume-limited subsample from NSA catalog (contour levels not linearly separated), the grey shading is MaNGA sample corrected by galaxy weights, and the blue dots are WR galaxies with S0 type WR galaxies highlighted with additional red circles. Two side panels show the histograms of S{\'e}rsic index for the aforementioned three samples, one panel for $M_*>10^{10}M_\odot$ for all three samples and the other for $M_*<10^{10}M_\odot$. Errorbars are Poisson counting error. NSA-based and MaNGA-corrected samples overlap at some bins, for they both reflect the general population.}
%    \label{sersic_mass2}
%\end{figure*}
% $n=6$ is the manual upper limit in fitting and this causes some pile-up at the top.

 In \autoref{structural} we show the distribution
  of structural parameters, comapring the WR galaxy sample with the
  MaNGA sample. For both samples we have weighted each galaxy using
  the weights provided from the MaNGA MPL-7 release. The structural
  parameters are measured by \citet{2016MNRAS.456.3032P} by applying a
  range of automated structural measures to the sky-subtracted
  $r$-band SDSS images. These include $n$ (S{\'e}rsic index to measure
  the power-law index in fitting the radial surface brightness
  profile), $G$ (Gini index to measure the degree of inequality in the
  light ditribution), $M_{20}$ (second-order moment of the
  flux-weighted distance of the brightnest pixels containing 20\% of
  the total light), $C$ (concentration index  defined by five times
  the logarithmic ratio of the radii enclusing 80\% and 20\% of the
  total light), $A$ (rotational asymmetry), and $S$
  (clumpiness). Definitions of these parameters are detailed in \S~3.2
  of \citet{2016MNRAS.456.3032P}.

  As can be seen, the general population of galaxies show clear
  dependence on mass in all the prameters, in the sense that  galaxies
  of lower masses have lower values of $n$, $G$ and $C$, higher values of
  $M_{20}$ and $S$, and similar values of $A$. In contrast, the WR
  galaxy sample appears to show no or weak mass dependence, with the
  two mass subsamples showing very similar distributions in all
  parameters. As a result, the WR galaxies share the same
  distributions of $n$, $M_{20}$ and $C$ as the general population of
  low-mass galaxies, while they are the same as the high-mass general
  population in terms of $G$ and $S$. At both masses the WR galaxies
  are more asymmetric than the general population, indicating a
  higher-than-average fraction of interacting/merging systems in the
  WR galaxy sample. In the rightmost panels we show the distributions
  of the morphological $T$-type again, but for the two mass ranges
  separately. At low masses the WR sample shows the same $T$-type
  distribution as the general population.  At high masses the WR
  galaxies are limited to late-type morphologies with few galaxies
  with $T<0$. Therefore, the WR galaxies of S0 type as previously
  seen in \autoref{morphology} are mostly at low masses.

In \autoref{gini-m20} we show the distribution of
   WR galaxies (blue dots) and general population for comparison (grey-scale
  background) on the plane of $G$ versus $M_{20}$. The general population is reconstructed from MaNGA MPL-7 sample with weights. The diagram is separate for the
two mass intervals divided at $10^{10}M_\odot$. The
higher $G$ of the WR sample at low masses and their lower $M_{20}$ at
high masses, as seen in the previous figure, are also clearly seen
here. As a result, the WR galaxies at both masses are preferentially
found in the area of ``merging systems'', as defined by the empirical
diagnostic lines from \citet{2008ApJ...672..177L} and
\citet{2004AJ....128..163L} which are plotted as the solid and dashed
lines in the figure.  This result is consistent with the
larger-than-average asymmetry of the WR sample as found in the
previous figure.

  To further understand the merging fraction of the WR sample,
  we have separated the WR galaxies in interacting systems from the
non-interacting WR galaxies using the pair galaxy catalog from
\citet{Fu-18}, who identified interacting and merging galaxies
  in MaNGA MPL5 and later extended to MaNGA MPL6 (basically identical to MPL7) by the criteria of ``projected separations less than 30 kpc, radial velocity offsets less than 600 km s$^{-1}$, and mass ratios greater than 0.1''. We find 43 out of the 90 WR galaxies
(47.8\%) to be interacting galaxies or mergers, a fraction that is
much higher than that of the general population. This strongly suggests
that the WR galaxies, as identified usually with strong star formation,
are closely associated with galaxy-galaxy interactions. This finding
is consistent with the known effect that tidal interactions between
galaxies can effectively enhance the star formation in galactic
centers \citep[e.g.][]{Li-08}.

\section{Discussion}
\label{sec:discussion}

\subsection{Comparison with SDSS-based catalogs}

\begin{figure*}[h!]
\centering
\includegraphics[width=0.95\textwidth]{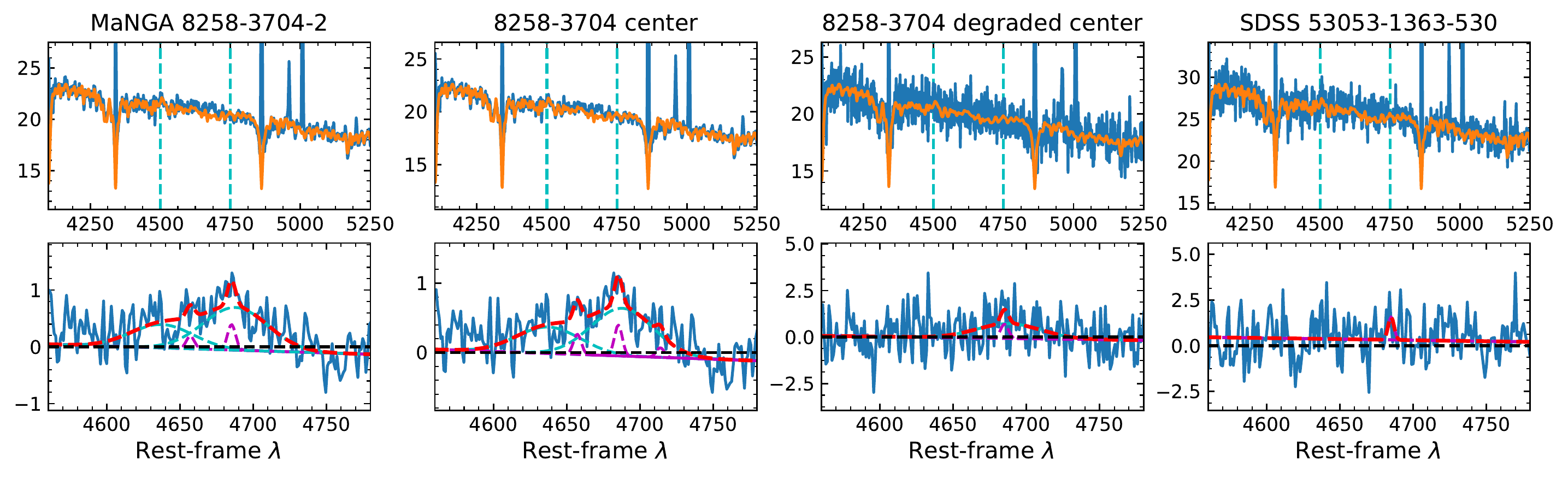}
\includegraphics[width=0.95\textwidth]{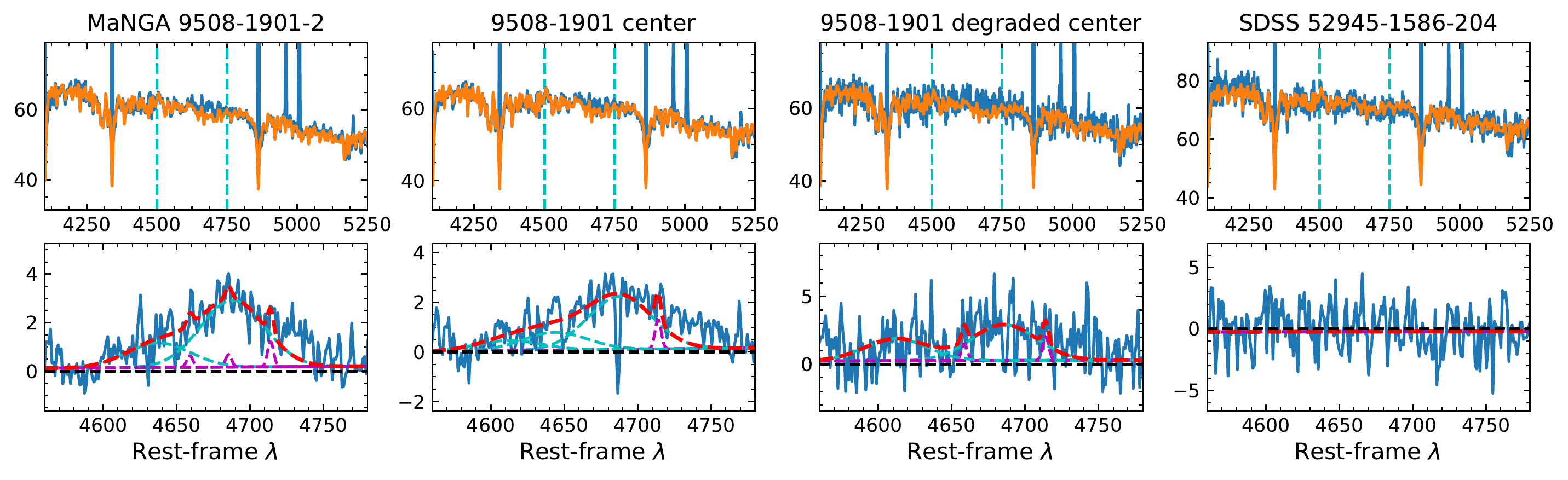}
\includegraphics[width=0.95\textwidth]{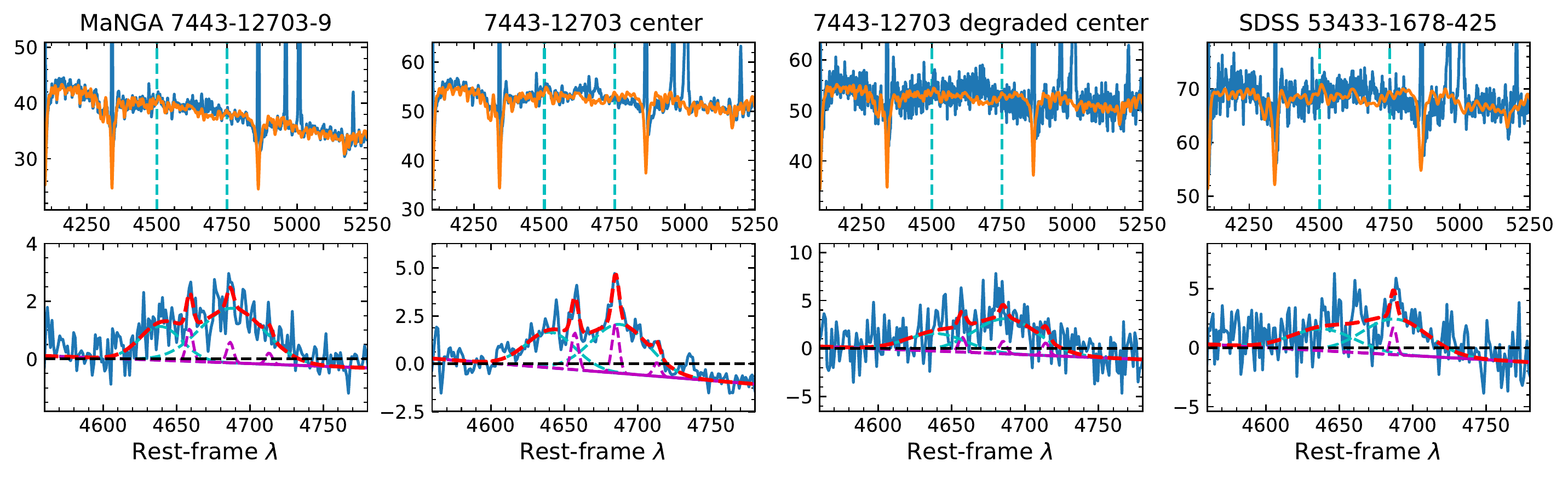}
\includegraphics[width=0.95\textwidth]{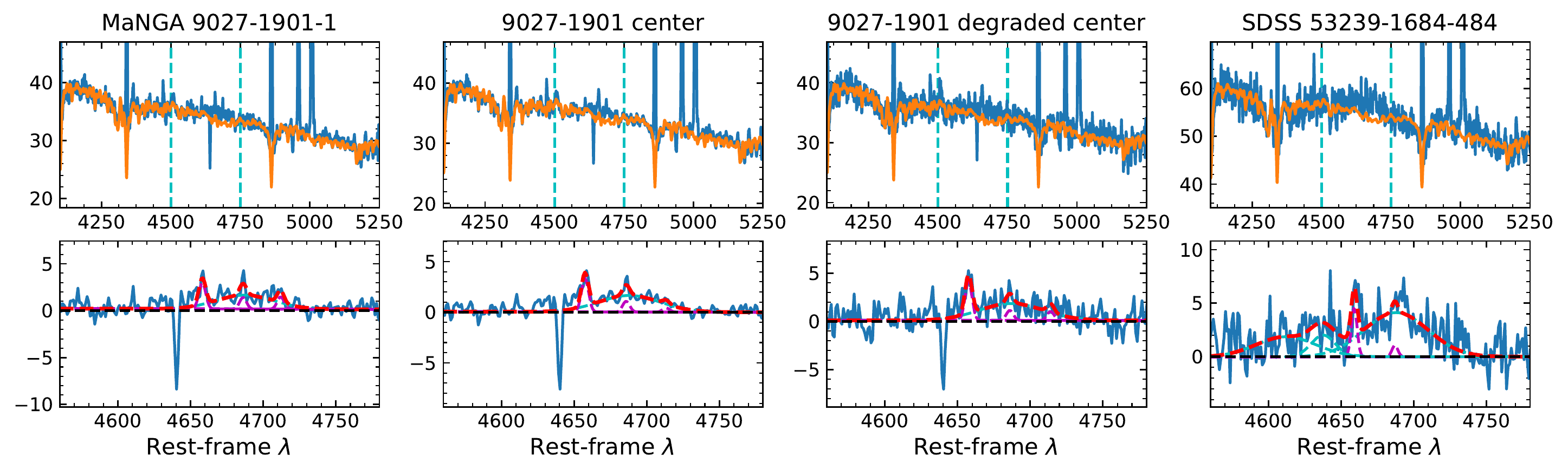}
\caption{Comparison with SDSS spectra. Four sets of panels are shown. Each set is one example galaxy of the following categories: WR galaxies missed by SDSS studies purely due to low spectral S/N, WR galaxies missed by SDSS studies due to a difference in fitting models (also related to low spectral S/N), WR galaxies showing WR features even in SDSS spectra but missed by SDSS studies due to unclear reasons, and one WR galaxy identified by SDSS studies but missed by this work. In each set, the top row shows the observed spectra and the respective fitting models for them, and the bottom row shows the residual spectra in the zoom-in WR wavelength range with a tentative WR bump fitting. The first column in each set shows the central WR region; the second column is the stacked central 3{\arcsec}-diameter spectrum; the third column is the degraded second column to match the S/N of the fourth column, which is the original SDSS spectra.}
\label{fig:sdss_comparison}
\end{figure*}

\citet{Zhang-07} selected 174 WR galaxies from SDSS data release 3
(DR3), while \citet{Brinchmann-Kunth-Durret-08} selected 570 WR
galaxies from SDSS data release 6 (DR6).  The fractions of WR galaxies
are 0.05\% and 0.08\% without correcting for incompleteness due to
flux limit or other selection effects. The SDSS observed only the
central 1-2 kpc of the galaxies using a single fiber of
3\arcsec\ diameter. Out of the 90 WR galaxies selected from the
MaNGA MPL-7 in this work, we find 46 galaxies to have a central WR
region, i.e. they have at least one WR region whose center spaxel is
located within the central 3\arcsec-diameter region.  Among these 46
galaxies, 39 are in SDSS DR6, whiles others are added from SDSS DR7 or
the SDSS-III BOSS project. Of the 39 galaxies, only 7 are included in
the catalog of \citet{Zhang-07} or \citet{Brinchmann-Kunth-Durret-08}.
In other words, the SDSS-based catalogs have missed most of the WR
galaxies from MaNGA (i.e. 32/39), even when we only consider the WR
regions in galactic centers. 

In order to understand why the SDSS-based studies failed to identify
these 32 galaxies, we have done a one-to-one comparison between
SDSS data and MaNGA data, and we find multiple reasons. 
Before we start the detailed comparison, for six galaxies, we find the SDSS
fiber was positioned on an off-center region, and so the central WR
region is outside the SDSS fiber but covered by the MaNGA
IFU. These six galaxies should not be counted as missing WR galaxies in previous studies.

In the remaining 26 comparisons, first of all, about 70\% of the missing galaxies (18/26) should be attributed
to the low S/N of the SDSS spectra. In \autoref{fig:sdss_comparison},
as an example, row No.1-2 show spectra and WR bumps of the central
WR region from MaNGA (left-most panel) and the SDSS spectrum of the
same galaxy (right-most panel). The upper panels show the observed
spectrum and the best-fit stellar spectrum from 4000-5250 {\AA}, and
the lower panels show the starlight-subtracted spectrum over the WR
bump wavelengths. We obtain the best-fit stellar spectrum for all the
cases using our fitting code (see \S~\ref{sec:fitting}).  We note that
the central WR region may cover different area than the SDSS fiber,
and the spectra from the two surveys have different S/N. To have a
more direct comparison, we have obtained the integrated spectrum over
the central 3\arcsec-diameter region (i.e. the same as covered by the
SDSS fiber) with our stacking code from the MaNGA datacube. The observed, best-fit and
residual spectra are shown in the second column in which the WR
feature is similarly seen. The third column shows the MaNGA spectra of the
central 3\arcsec-diameter region again, but random noise is added to
the observed spectrum so as to have the same S/N as the SDSS
spectrum. As can be seen, like the SDSS spectrum, the WR bump becomes
very weak and the region would be unlikely to be identified as a WR
region. 

For the remaining eight galaxies, we find five galaxies can also be
attributed to their low spectral S/N, but in a different manner ---
they were missed due to uncertainties in the spectral fitting of low
quality spectra. Row No. 3-4 of \autoref{fig:sdss_comparison} shows
an example galaxy in this case. For this galaxy, the WR bump is still
significant (though weaker) after the S/N of the spectrum is reduced
to match the S/N of the SDSS spectrum. However, the residual spectrum
of the SDSS shows almost no feature in the WR wavelength
window. Looking closely at the upper panels in the third and the last
column of the second row, we find the best-fit stellar spectra indeed
differ slightly. This indicates that the identification of WR features
could be affected by spectral fitting especially when the S/N of the
spectrum is relatively low. Finally, for the last three galaxies, we find their WR feature
remains after the S/N is reduced, and this is true also for the SDSS
spectrum. Row No. 5-6 of the figure displays one of the three
galaxies as an example. They were missed by previous studies possibly
because of different criteria in visual inspection, their different fitting recipes or other
serendipitous reasons.

We conclude that the much lower detection rate of WR galaxies in the
SDSS is caused by multiple reasons. Among all reasons, the limited
spatial coverage of the single-fiber spectroscopy and the relatively
low spectral S/N are the main reasons, which can respectively explain
about a half and 70\% in the remaining half of WR galaxies that were missed in
previous SDSS-based studies.

We should point out that, there is only one galaxy which was
identified to be a WR galaxy in SDSS-based studies but is not included
in our WR galaxy catalog. The observed, best-fit and residual spectra
of this galaxy are shown in the bottom two rows of
\autoref{fig:sdss_comparison}. As can be seen, the galaxy presents a
blue bump in all cases, indeed, although it appears to be more
pronounced in the SDSS spectrum. This galaxy is missed in our case,
likely due to the relatively strict criterion of our visual
inspection. We tend not to add this galaxy to our catalog to keep the
consistency of our selection procedure. We should keep in mind,
however, that the detection fraction of WR galaxies in our study
should be regarded as a lower limit, although it is much closer to the
real fraction when compared to the SDSS catalogs. 

\subsection{Comparison with CALIFA-based catalog}

Although the MaNGA sample gives a much higher detection rate than
previous SDSS samples, the WR galaxy fraction ($1-2$\% at most stellar
masses, see \autoref{fraction-mass}) is  a bit too low when compared
to the CALIFA sample which includes a fraction of $\sim$4.5\% WR
galaxies \citep{Miralles-Caballero-16}. This difference can be
explored from a couple of factors. The first is the different
spatial coverages of the CALIFA and MaNGA IFUs. Although CALIFA
covers galaxies out to their 3-4 $R_e$, which is larger than the
1.5$R_e$ and 2.5$R_e$ covered by the MaNGA Primary and Secondary
samples, we find that all WR regions in CALIFA galaxies are located
within 1.5 $R_e$. Therefore, the spatial coverage of the IFUs should
not cause any difference in the WR galaxy fraction. Secondly, we
consider the possible effect of the different spatial resolutions. The
angular resolutions are {2.5\arcsec} for both surveys \citep{Law-16,
  2015-CALIFA-DR2}.  Given the different redshifts and field of views of the two surveys,
the physical resolution (in unit of kpc) and the relative resolution
(in unit of $R_e$) could both be different. We discuss these two
factors below.

\begin{figure}
	\includegraphics[width=\columnwidth]{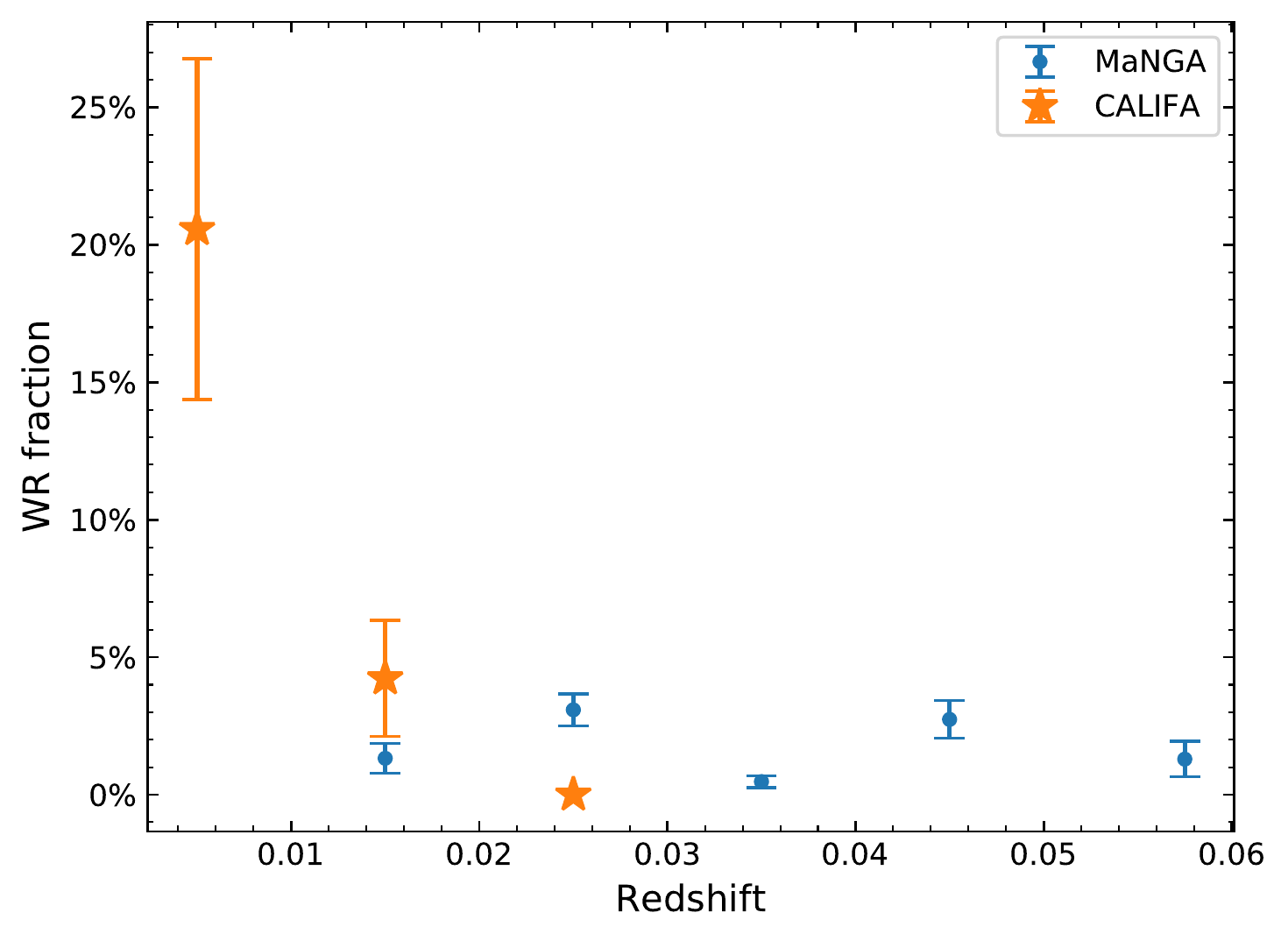}
    \caption{WR fraction as a function of redshift for both MaNGA  (blue dots) and CALIFA (orange stars) catalogs. Both fractions have been applied the volume correction.}
    \label{frac-z}
\end{figure}

The physical resolution is mainly determined by redshift given the
same angular resolution. We find that most CALIFA WR galaxies
(11 out of 15 from the main sample, 20 out of 25 from the entire
sample) are below $z=0.01$ while MaNGA galaxy sample has a lower limit
of $z=0.01$.  The CALIFA parent sample has an approximately uniform
distribution from $z=0$ to $z=0.03$. Therefore, it is apparent that
when we limit the redshift range to $0.01<z<0.03$, CALIFA detection
rate becomes pretty low. As for MaNGA, the
volume-corrected detection rate in $0.01<z<0.03$ increase to 2.28\%,  due to a better
physical resolution at lower redshift. In z=0.01-0.02, see \autoref{frac-z}, MaNGA and CALIFA have very similar WR fraction. In z=0.02-0.03, CALIFA has no WR galaxy detected, maybe due to Poisson fluctuation with its relatively small sample size. The much higher detection rate from CALIFA at $z<0.01$ may imply the real WR galaxy rate is still higher than the fraction of CALIFA and MaNGA studies. 
Overall, due to small
number statistics, we are not sure whether the high overall fraction
of CALIFA WR is entirely due to the contribution from $z=0-0.01$.

Then we consider the relative resolution. With the CALIFA
field of view  and survey design, the angular diameter of its galaxies
are typically 2-4 times larger than MaNGA galaxies. So with the same
angular resolution, CALIFA sample have a better relative resolution
with regard to $R_e$. Since the overlap of galaxy size between CALIFA
and MaNGA is pretty small, we can hardly construct subsamples from the two surveys with the same distribution, and thus we resort to another approach for testing
this factor. 
We use MaNGA sample to examine whether the detection rate has a trend with galaxy size. Size is correlated with other basic properties over the process of galaxy evolution, for example mass, which significantly affects WR rate as shown previously. Thus, we control mass when examining the correlation of WR rate with size, as mass is the most driving dependency of WR rate among basic properties of galaxies. Our result shows a slight decrease of detection rate towards larger galaxy size, on the contrary to the higher detection rate of CALIFA with its larger galaxy sizes. Since there is large uncertainty on the trend due to small number of statistics, we limit our discussion to this phenomenological trend and leave the validation of the trend as well as physical interpretation to future studies.
%We divide MaNGA sample into different size bins and
%examine whether the detection rate has a trend with galaxy
%size. Because size itself is dependent on mass and WR detection rate
%varies significantly with mass, we control mass when comparing WR detection rate with size.
%Generally we see a slight decrease of detection rate towards larger galaxy
%sizes, on the contrary to the higher detection rate of CALIFA with its
%larger galaxy sizes. Since galaxy size is also dependent on multiple other parameters and there is large uncertainty on the trend,
%we limit our discussion to this phenomenological trend alone and leave the physical interpretation to future studies.

Next, we consider the difference in parameter distributions of galaxies in the two surveys which may also contribute to the different detection
rates. Both CALIFA and MaNGA consist of a main sample and some
ancillary programs. The ancillary programs mostly target peculiar
galaxies and therefore increase the uncertainty and difficulty of
constraining the WR fraction. For this reason, we limit our discussion
here to WR galaxy fraction in main samples of the two surveys.  For
CALIFA we discard the extension sample, and for MaNGA, we keep the
primary sample and secondary sample while excluding the color-enhanced
sample and galaxies from ancillary programs. For the CALIFA sample, 15 out of
the 25 WR galaxies are from its 448 main sample galaxies
\citep{Miralles-Caballero-16}.  In the case of MaNGA, 60 out of the 90 WR
galaxies are from its 3735 unique Primary or Secondary sample
galaxies. Without any corrections, the WR galaxy fraction is 3.35\%
and 1.61\% for CALIFA and MaNGA, respectively. Considering the mass
dependence of the WR galalaxy detection as shown in
\autoref{mass-function}, we construct a control sample of galaxies
from the MaNGA Primary and Secondary samples that have the same
distribution in stellar mass as the CALIFA main sample. We find a
WR galaxy fraction of $1.26\pm 0.14$\% in this control sample.
Furthermore, we require the control sample of MaNGA galaxies to have
the same distribution as CALIFA in both stellar mass and $NUV-r$
color, to take into account potential effect of different colors
of the two samples. To this end, we have trimmed both samples so that
they have similar distributions of galaxies in the two-dimensional
mass---color space. The resulting WR galaxy fractions are
$1.63 \pm 0.24 \%$ for MaNGA and $3.68 \pm 0.95\%$ for CALIFA. The errors are Poisson counting error. 
Therefore, we conclude that the different WR galaxy detection rates between the two surveys
are not caused by the different properties of the galaxies.

Also, MaNGA and CALIFA spectra have similar S/N distribution.
Therefore, S/N should not be a significant reason for the difference
in WR fraction.

Finally, we notice that the WR galaxy catalogs from both CALIFA and
MaNGA have a large fraction of merging systems. We use \citet{2014A&A...569A...1W} for classification of CALIFA galaxies and 
the extended version of \citet{Fu-18} for classification of MaNGA galaxies.
By separating mergers and isolated galaxies, we find the WR fraction in isolated galaxies is $3.15\pm 0.8\%$ for CALIFA and $1.41\pm 0.20$\% for MaNGA. Errors are Poisson counting errors. The difference is still significant. 
Therefore, we claim the difference in WR fractions of the two surveys is unlikely to be caused by the merger fraction in their parent samples.
But we also want to point out that due to the small numbers in statistics and the subjectivity in classification of mergers, 
we still need better data and classification to explore the effect of mergers on WR fraction in the future.

To conclude, the higher WR galaxy fraction in the CALIFA sample may be
explained mainly by the inclusion of galaxies at $z<0.01$. Above the MaNGA redshift
limit $z=0.01$, the WR galaxy fractions from the two surveys are
actually very similar. Other factors, whether physical or instrumental, do not have a clear or significant contribution in the 
difference between this work and CALIFA WR catalog.

\subsection{Abundance of WR galaxies}
\label{fraction}

Throughout the construction of this catalog, we put much emphasis on
purity, especially in the careful choice of full spectrum fitting
recipe and visual inspection. As for the completeness, we suspect that
there should always be some weak WR population that is beneath the
detection capability of our data quality. For example, in the
determination of \ion{H}{2} regions, we adopt a high threshold for
$H\alpha$ surface brightness, which may possibly miss some WR
regions. Therefore, our WR fraction of $1.9 \pm 0.2 \%$ should be
considered as \textit{lower limits} of the real fractions. The higher
WR detection rate of CALIFA, as discussed above, also indicates that
the WR galaxy fraction could be even higher if IFU surveys with higher
resolution and S/N are available.

\subsection{Coexistence of WR features and Active Galactic Nuclei}
\label{wr-agn}

There are three WR galaxies in our catalog that show either AGN-like
broad emission lines or Seyfert line ratios on the BPT diagram
\citep{Baldwin-Phillips-Terlevich-81, Kewley-01, Kauffmann-03,
  Kewley-06}. For example, the outlier at redshift z$\simeq$0.11 in
\autoref{mass-z} is one of them. We carefully examined the spectra,
spatially-resolved BPT diagram, Fe emission line template from
\citep{Veron-Cetty-Joly-Veron-04}, etc. We conclude all these galaxies
show real WR features with possible simultaneous existence of AGN. For
example, galaxy 8626-12704 has a clear WR red bump. We also find a few
WR regions classified as "composite of star-forming and AGN
activities" on the BPT diagram and these galaxies may carry specific
scientific interest. With the coexistence of WR population and AGN, it
is possible to study the interaction between AGN and recent
star-formation in the future.

\subsection{Red bump}
As mentioned earlier, besides the blue bump, WR stars also form a red
bump around 5800 {\AA} with C and N broad emission lines. Normally the red bump
feature is weaker than the blue bump. With our WR catalog, we visually
examine the wavelength range of the red bump and find 39 WR regions
from 24 WR galaxies show red bumps. The weakness of the red bump is not only reflected in the small number of occurrence but also in the significance of individual occurrences, and therefore these identifications carry higher uncertainty compared to the blue bump identifications.
With these red bumps, together with information from the blue bumps, we can further
derive the ratio among different WR subtypes. We will leave this part
to the parallel paper discussing spatially resolved properties of WR
regions. % (Liang et al. in prep)

\section{Conclusions}
\label{sec:conclusion}

In this work we have carried out a thorough search for WR galaxies
from MaNGA MPL7 (i.e. SDSS DR15) data. We develop a two-step searching
scheme to tackle the challenge of the weakness of WR features.
We start by identifying \ion{H}{2} regions in the MaNGA datacubes.
We obtain a high S/N spectrum for each region by stacking the
original spectra, and perform full spectral fitting to the stacked
spectrum. Next, we visually examine the starlight-subtracted spectrum
of all the \ion{H}{2} regions, and identify WR regions according
to the presence of a blue bump at $4600-4750$ {\AA} as signature of
the WR stars. The resulting WR catalog consists of 267 WR regions,
distributed in 90 WR galaxies, which is 1.9\% of the parent
sample. This fraction is much higher than previous stuidies
based on single fiber SDSS data, and similar to the recent
study based on CALIFA IFU data. Through detailed comparisons with
SDSS and CALIFA surveys, we evaluate the impact of different
survey parameters on WR fraction, and show consistency between this
work and previous studies. We have examined the global properties
of the WR galaxies, and for the first time estimated the stellar
mass function of WR galaxies.

Our main conclusions can be summarized as follows.
\begin{itemize}
  \item WR regions are exclusively found in galaxies that show bluest
    colors and highest star formation rates for their mass, as well as
    late-type dominated morphologies. 
    Their images have relatively high asymmetry, indicative of a higher-than-average
    fraction of mergers in the WR population.
  \item The stellar mass function of WR galaxies can well be described
    by a Schechter function with amplitude $\phi_*$ = 0.000157 Mpc$^{-3}$, 
    characteristic mass $log_{10}(M^*/M_\odot)$ = 10.332, and the faint-end slope $\alpha$ =
-0.905. This gives  rise to an average number density of $3.47\times 10^{-4}Mpc^{-3}$ and an average detection
    rate of 1.4\% with respect to the general population of galaxies in
    the Local Universe. The detection rate shows weak dependence on stellar mass,
     with a maximum of $\sim4$\% at $M_\ast\sim10^{9.7}M_\odot$.
  \item The small fraction of WR galaxies found previously in SDSS-based
    samples is attributed mainly to two facts. One is the single-fiber
    spectroscopy covering a limited central region of galaxies. The second
    fact is the lower S/N of the SDSS spectra compared to MaNGA.
    About half of our WR galaxies show WR features in their centers, but
    most of them were missed by previous SDSS studies due to the low
    S/N of the SDSS spectra.
  \item The CALIFA finds a higher fraction of WR galaxies than MaNGA
    mainly due to the inclusion of galaxies at $z<0.01$, which have
     better spatial resolution (in unit of pc) than galaxies at higher redshift.

\end{itemize}

There are still some limitations of this study for future
improvements. Although MaNGA has its unique advantage of a large
sample size, allowing us to construct a large catalog of new WR
galaxies, its S/N and spatial resolution are not ideal for
WR search. Random error of spectra may cause false positive
identification in a few cases while the current spatial resolution may
lead to loss of some compact WR regions due to dilution
effect. Furthermore, the weak WR feature is very sensitive to full
spectrum fitting recipe. The flux calibration of fitting templates,
the masking of emission lines, the fitting code and fitting procedure
in use, etc. may all affect the WR feature in the fitting residual. In
the future, better templates such as SSPs derived from MaStar stellar
library \citep{2010ApJS..186..427N} and deeper exposure will probably
improve this study.  Nevertheless, our catalog includes a large number
of WR regions from galaxies covering wide ranges in mass and color,
and so it should be able to form a good basis for many future studies.
In fact, the current paper is the first of a series of works in which
we will perform extensive studies of the WR regions/galaxies. The next paper will include resolved
mass-metallicity relation of WR regions. Also, WR subtype ratios will be studied as a function of metallicity and stellar population age. Moreover, both WR blue bumps and red bumps will be compared to model predictions from Starburst99 \citep{1998ApJ...497..618S} and BPASS \citep{2017PASA...34...58E} to constrain the metallicity-dependent variation of the stellar
initial mass function.   % (Liang et al. in prep.)

\acknowledgments
We are grateful to the anonymous referee for his/her helpful comments.
This work is supported by the National Key R\&D Program of China
(grant Nos. 2018YFA0404502) and the National Science Foundation
of China (grant Nos. 11821303, 11973030, 11761131004, 11761141012, and 11603075).

Funding for SDSS-IV has been provided by the Alfred P. Sloan Foundation and Participating Institutions. Additional funding towards SDSS-IV has been provided by the US Department of Energy Office of Science. SDSS-IV acknowledges support and resources from the Centre for High-Performance Computing at the University of Utah. The SDSS web site is www.sdss.org.

SDSS-IV is managed by the Astrophysical Research Consortium for the Participating Institutions of the SDSS Collaboration including the Brazilian Participation Group, the Carnegie Institution for Science, Carnegie Mellon University, the Chilean Participation Group, the French Participation Group, Harvard–Smithsonian Center for Astrophysics, Instituto de Astrofsica de Canarias, The Johns Hopkins University, Kavli Institute for the Physics and Mathematics of the Universe (IPMU)/University of Tokyo, Lawrence Berkeley National Laboratory, Leibniz Institut fur Astrophysik Potsdam (AIP), Max-Planck-Institut fur Astronomie (MPIA Hei- delberg), Max-Planck-Institut fur Astrophysik (MPA Garching), Max-Planck-Institut fur Extraterrestrische Physik (MPE), National Astronomical Observatory of China, New Mexico State University, New York University, University of Notre Dame, Observatario Nacional/MCTI, The Ohio State University, Pennsylvania State University, Shanghai Astronomical Observatory, United Kingdom Participation Group, Universidad Nacional Autonoma de Mexico, University of Arizona, University of Colorado Boulder, University of Oxford, University of Portsmouth, University of Utah, University of Virginia, University of Washington, University of Wisconsin, Vanderbilt University and Yale University.

A. Roman-Lopes acknowledges financial support provided in Chile by Comisi\'on Nacional de Investigaci\'on Cient\'ifica y Tecnol\'ogica (CONICYT) through the FONDECYT project 1170476 and by the QUIMAL project 130001

\bibliography{wr,ref}
\bibliographystyle{aasjournal}

\end{document}